\documentclass[12pt]{article}

\baselineskip=\normalbaselineskip

\usepackage{mathrsfs,amsbsy,amssymb,latexsym,amsfonts,amsmath,amsthm}
\usepackage{graphicx,color}

\allowdisplaybreaks[2]
\renewcommand{\thefootnote}{\arabic{footnote}}
\makeatletter
\catcode`\@=11
\@addtoreset{equation}{section}

\def\@seccntformat#1{\csname the#1\endcsname.~~}
\makeatother

\newcommand{\nn}{\nonumber}

\begin{document}

\begin{titlepage}
\renewcommand{\thefootnote}{\fnsymbol{footnote}}
\begin{flushright}
KUNS-2547
\end{flushright}
\vspace*{1.0cm}

\begin{center}
{\Large \bf 
Random volumes from matrices
}
\vspace{1.0cm}

\centerline{
{Masafumi Fukuma,}%
\footnote{E-mail address: 
fukuma\_at\_gauge.scphys.kyoto-u.ac.jp} 
{Sotaro Sugishita}%
\footnote{E-mail address: 
sotaro\_at\_gauge.scphys.kyoto-u.ac.jp} and
{Naoya Umeda}%
\footnote{E-mail address: 
n\_umeda\_at\_gauge.scphys.kyoto-u.ac.jp}%
}

\vskip 0.8cm
{\it Department of Physics, Kyoto University, Kyoto 606-8502, Japan}
\vskip 1.2cm 

\end{center}

\begin{abstract}

We propose a class of models which generate three-dimensional random volumes, 
where each configuration consists of triangles glued together along multiple hinges. 
The models have matrices as the dynamical variables 
and are characterized by semisimple associative algebras $\mathcal{A}$. 
Although most of the diagrams represent configurations which are not manifolds, 
we show that the set of possible diagrams can be drastically reduced 
such that only (and all of the) three-dimensional manifolds 
with tetrahedral decompositions appear,  
by introducing a color structure 
and taking an appropriate large $N$ limit.
We examine the analytic properties when $\mathcal{A}$ is a matrix ring or a group ring, 
and show that the models with matrix ring have a novel strong-weak duality 
which interchanges the roles of triangles and hinges. 
We also give a brief comment on the relationship of our models 
with the colored tensor models. 

\end{abstract}
\end{titlepage}

\pagestyle{empty}
\pagestyle{plain}

\tableofcontents
\setcounter{footnote}{0}
\section{Introduction}
\label{intro}

String theory is a strong candidate for a unified theory including quantum gravity. 
However, it still does not have a constructive, nonperturbative definition. 
The main reason is the lack of our understanding 
on the real fundamental dynamical variables of string theory. 
In fact, 
since the advent of D-branes and the discovery of string dualities, 
the idea has widely spread 
that the fundamental dynamical variables  
need not be strings and can be other types of extended objects.

M-theory \cite{Witten:1995ex} is a description of string theory, 
where membranes are believed to play an important role.%
\footnote{
The BFSS matrix model \cite{Banks:1996vh} is another candidate 
of nonperturbative definition of M-theory, 
where D0-branes play the fundamental roles 
(see \cite{Taylor:2001vb} for a review). 
} 
The worldvolume theory of membranes is equivalent to a three-dimensional gravity theory, 
where the target space coordinates of an embedded membrane 
are expressed as scalar fields in three-dimensional worldvolume 
(see \cite{Taylor:2001vb} for a review). 
However, the analytic understanding of three-dimensional quantum gravity 
is still not sufficient, 
as compared to that of two-dimensional quantum gravity 
\cite{Knizhnik:1988ak, David:1988hj, Distler:1988jt}.

Here, the roles played by matrix models in string theory should be suggestive,  
where the Feynman diagrams of matrix models are interpreted as 
triangular (or polygonal) discretization of string worldsheets 
(see \cite{Ginsparg:1993is, Di Francesco:1993nw} for reviews). 
Furthermore, 
by introducing the degrees of freedom corresponding to matters on worldsheets  
or by considering a matrix field theory,  
one can define various kinds of string theory in terms of matrix models 
\cite{Kazakov:1988ch}.  
The $1/N$ expansion of matrix models, where $N$ is the size of matrix, 
corresponds to  the genus expansion of string worldsheets as in \cite{'tHooft:1973jz}. 
Moreover, the double scaling limit enables us to study 
the nonperturbative aspects of string theory 
\cite{Brezin:1990rb, Douglas:1989ve, Gross:1989vs, 
Fukuma:1990jw, Dijkgraaf:1990rs, Shenker:1990uf}  
as well as their integrable structure 
\cite{Douglas:1989dd, Fukuma:1990yk, Fukuma:1991ky}.

Tensor models \cite{Ambjorn:1990ge, Sasakura:1990fs, Gross:1991hx} 
or group field theory \cite{Boulatov:1992vp, Freidel:2005qe}
are natural generalizations of matrix models 
to three (and higher) dimensions.
For three-dimensional models,    
the perturbative expansion generates random tetrahedral decompositions 
of three-dimensional objects.  
Unlike the two-dimensional case, however, these objects 
are not always manifolds 
or not even pseudomanifolds.   
Recently the situation was drastically improved by the colored tensor models 
(see, e.g.,  \cite{Gurau:2011xp} for a review). 
It is shown that the colored tensor models admit a large $N$ expansion 
and the leading contributions represent higher dimensional sphere 
\cite{Gurau:2011xq, Bonzom:2012hw}.  
Moreover, it is claimed that 
one can take a double scaling limit in the tensor models 
\cite{Dartois:2013sra, Bonzom:2014oua}. 
Thus, the colored tensor models give a fascinating formulation 
of higher dimensional quantum gravity.  
Nevertheless,  the analytic treatment of tensor models is still not so easy 
as that of matrix models. 
For example, tensors cannot be diagonalized as matrices can, 
and an analogue of saddle point method has not been found yet.

In the present paper, 
we propose a new class of models which generate three-dimensional random volumes,  
by regarding each random diagram 
as a collection of triangles glued together along multiple hinges 
as in \cite{Chung:1993xr}.%
\footnote{
We confine our attention to three-dimensional pure gravity. 
The inclusion of matters will be discussed in our future communication. 
} 
Our models have real symmetric matrices as the dynamical variables 
and are characterized by semisimple associative algebras $\mathcal{A}$. 
Although most of the diagrams represent configurations which are not manifolds,%
\footnote{
In this paper, by a manifold we always mean a closed combinatorial manifold, 
which is a collection of tetrahedra 
whose faces are identified pairwise 
and each of whose vertices has a neighborhood 
homeomorphic to three-dimensional ball $B^3$. 
See, e.g., \cite{Ambjorn:1997di} for the rigorous definition.
} 
we show that the set of possible diagrams can be drastically reduced 
such that only (and all of the) three-dimensional  manifolds 
with tetrahedral decompositions appear,  
by introducing a color structure 
and taking an appropriate limit of parameters existing in the models.

Since our models are written with matrices, 
there should be a chance that various techniques in matrix models 
can be applied 
and the dynamics of random volumes can be understood more analytically.  
We show that our models have a novel strong-weak duality 
which interchanges the roles of triangles and hinges 
when $\mathcal{A}$ is a matrix ring.
This duality may suggest the analytic solvability of the models.

This paper is organized as  follows.   
In section \ref{def_of_model},  
we first define our models 
and show that the models are characterized by semisimple associative algebras 
$\mathcal{A}$. 
We then give a few examples of the Feynman diagrams, 
and show that some diagrams are not manifolds. 
From the examples, 
we deduce a strategy to restrict the models 
so that only (and all of the)  three-dimensional manifolds are generated. 
This strategy is implemented in section \ref{hm2mr}, 
where matrix rings are taken as the defining associative algebras. 
We explicitly construct models  
that generate only manifolds as Feynman diagrams, 
by introducing a color structure to the models 
and letting the associative algebras have centers 
whose dimensions play the role of free parameters. 
In section \ref{hm2gr}
we investigate the models 
where $\mathcal{A}$ is set to be a group ring $\mathbb{R}[G]$, 
and demonstrate how the models depend on details of the group structure of $G$. 
Section \ref{conclusion} is devoted to conclusion and discussions. 
We list some of the future directions for further study of the models, 
and give a brief comment on the relationship of our models 
with the colored tensor models.

\section{The models} 
\label{def_of_model}

In this section we define a class of models 
which have matrices as the dynamical variables 
and generate Feynman diagrams 
consisting of triangles glued together along multiple hinges. 
We show that the models can be defined by semisimple associative algebras. 
We then give a few examples of the Feynman diagrams, 
and show that some diagrams are not manifolds. 
We will conclude the section by giving a strategy to restrict the models 
so that only three-dimensional manifolds are generated. 
This strategy will be implemented in the next section, 
where matrix rings are taken as the defining associative algebras. 

\subsection{General structure} 
\label{subsec:general}

We first explain the diagrams we are concerned with 
and give the rule to assign a Boltzmann weight to each diagram. 
We then write down the action which generates such diagrams 
as Feynman diagrams.

\begin{figure}[htbp]
\begin{center}
\includegraphics[height = 3.5cm]{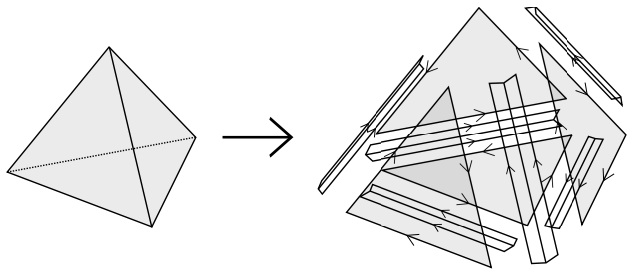}
\begin{quote}
\vspace{-5mm}
\caption{
Decomposition of a three-dimensional diagram 
to triangles and hinges. 
}
\label{fig:decomposition}
\end{quote}
\end{center}
\vspace{-6ex}
\end{figure}

\begin{figure}[htbp]
\begin{center}
\includegraphics[height = 2.0cm]{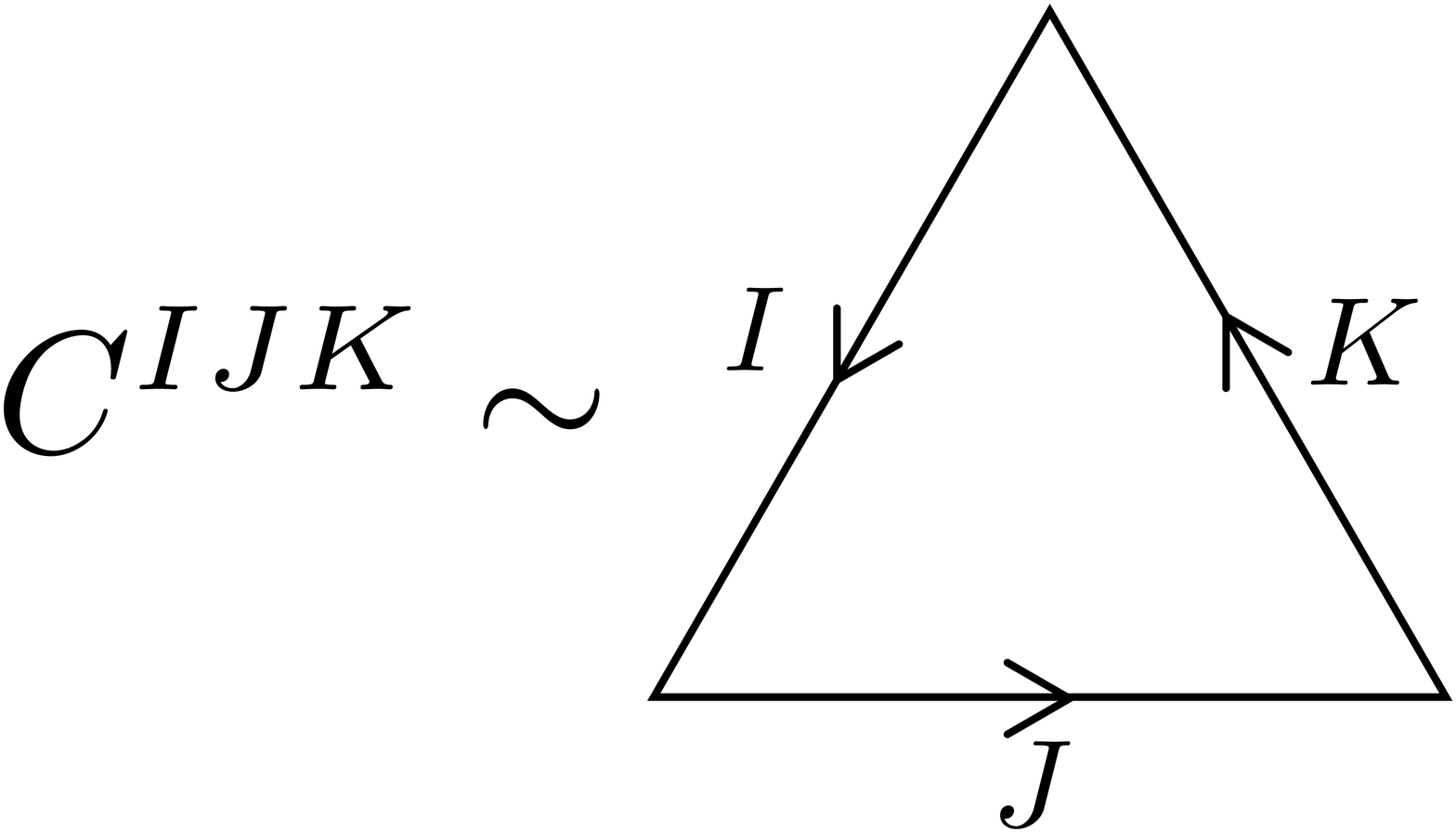}
\hspace{2cm}
\includegraphics[height = 2.5cm]{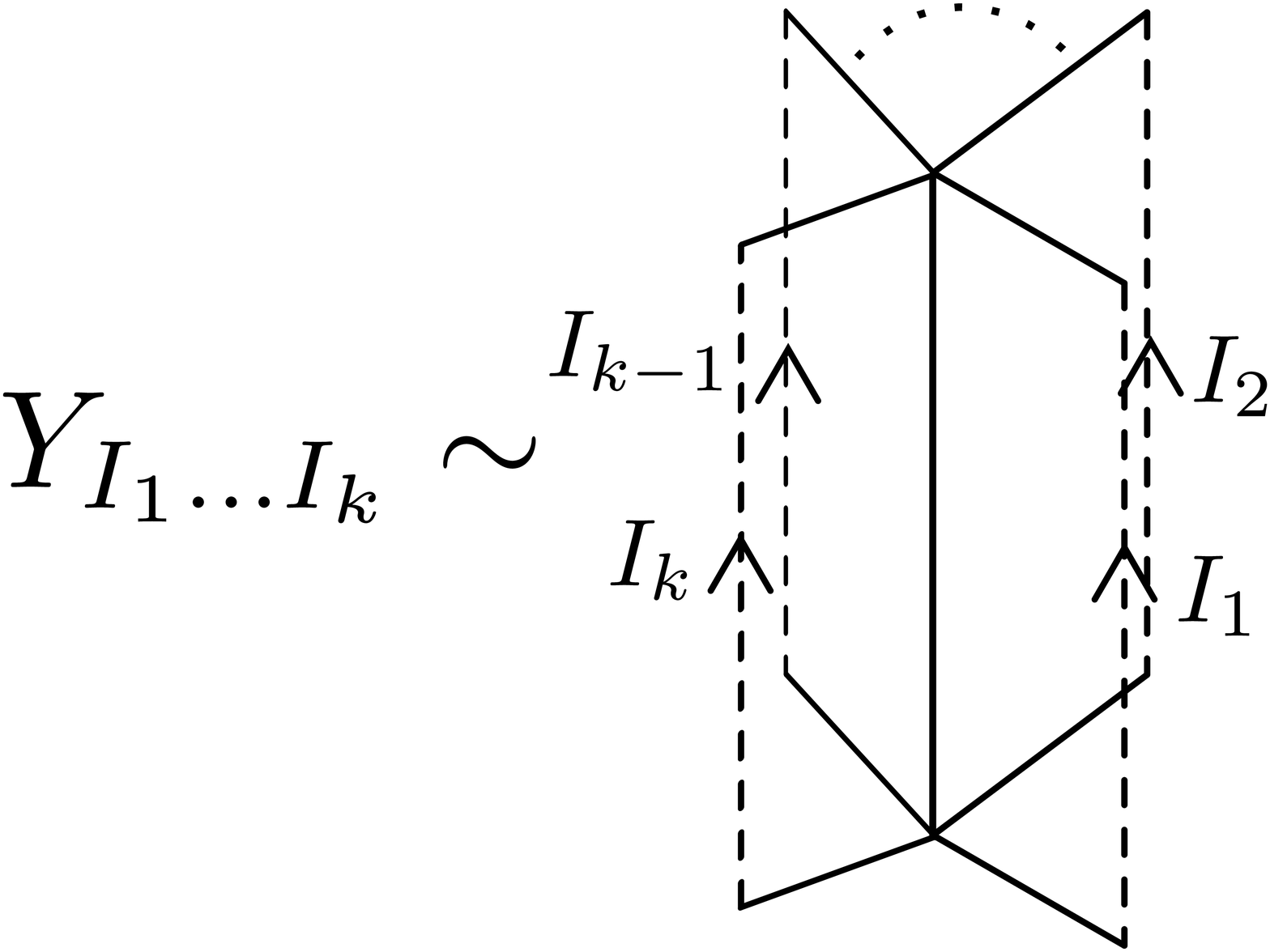}
\begin{quote}
\vspace{-5mm}
\caption{
Triangles and hinges. 
}
\label{fig:tri-hinge}
\end{quote}
\end{center}
\vspace{-6ex}
\end{figure}

We consider a set of diagrams, $\{\gamma\}$,  
consisting of triangles glued together 
along multiple hinges as in \cite{Chung:1993xr}. 
In order to assign a Boltzmann weight to diagram $\gamma$, 
we first decompose $\gamma$ 
to a set of triangles and a set of multiple hinges 
(see Fig.~\ref{fig:decomposition}). 
For each edge of a triangle, 
we draw an arrow 
and assign an index from a finite set $\{I\}$. 
We repeat the same procedure for the hinges. 
We then assign the real numbers $C^{IJK}$ and $Y_{I_1 \ldots I_k}$ 
to the indexed triangles and hinges, respectively,
as in Fig.~\ref{fig:tri-hinge}.%
\footnote{
The edges of a triangle will be drawn in solid lines 
while those of a hinge in dotted lines.  
} 
We require that $C^{IJK}$ and $Y_{I_1 \ldots I_k}$ be cyclically symmetric.

\begin{figure}[htbp]
\begin{center}
\includegraphics[height = 2.5cm]{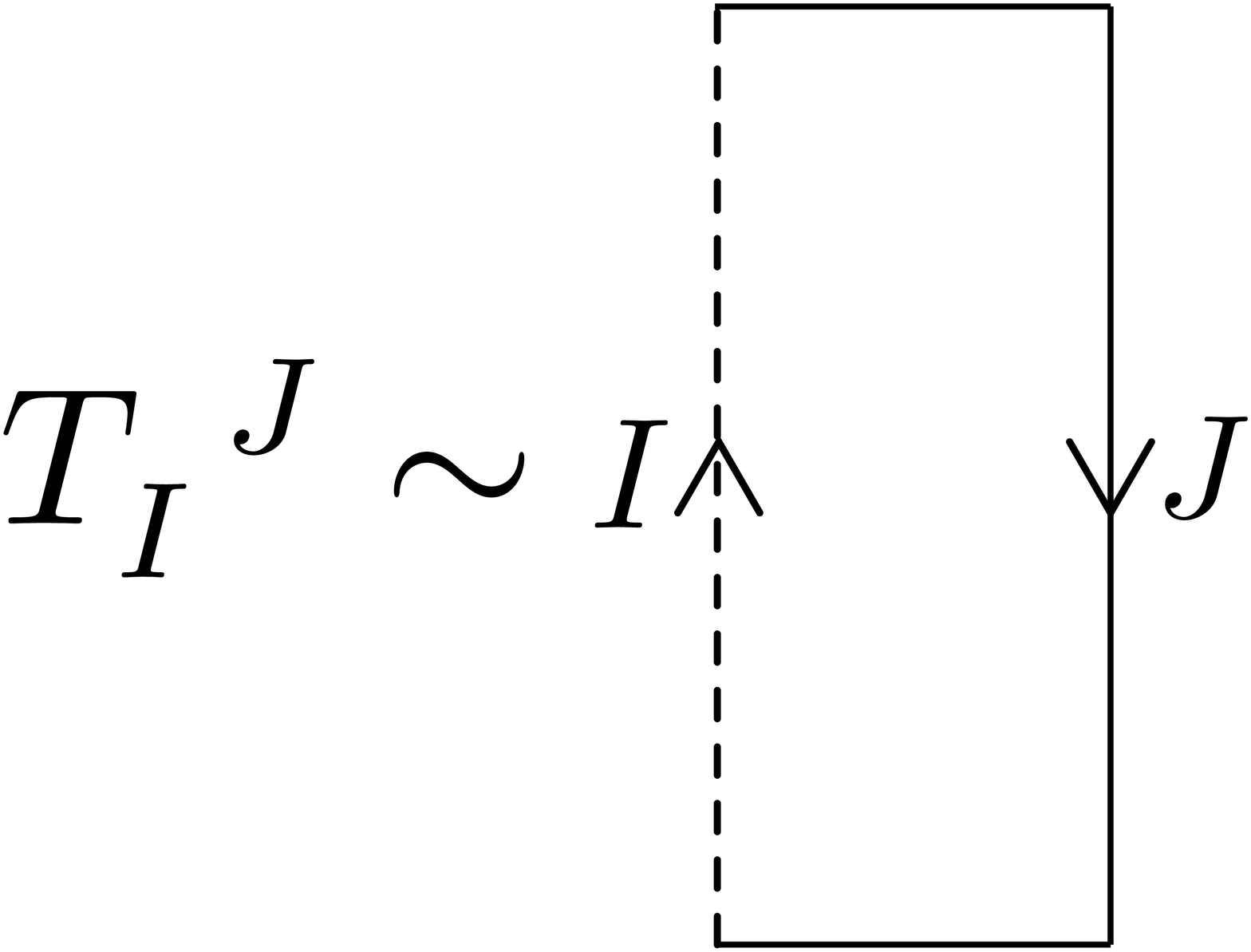}
\begin{quote}
\vspace{-5mm}
\caption{
Tensor $T_I^{\,\,J}$.  It changes the direction of arrow. 
}
\label{fig:T-tensor}
\end{quote}
\end{center}
\vspace{-6ex}
\end{figure}

Then, we glue the triangles and hinges 
to reconstruct the original diagram 
in such a way that the identified edges have the same index.  
In doing this, 
there may appear the case 
where the arrows of a triangle and a hinge have opposite directions.  
To treat such cases, 
we introduce a tensor $T_I^{\phantom{I}J}$ 
which reverses the direction of an arrow (see Fig.~\ref{fig:T-tensor}). 
Note that the tensor $T=(T_I^{\phantom{I}J})$ should be involutory 
because the direction of an arrow comes back to the original one  
after $T_I^{\phantom{I}J}$ is applied twice: 
\begin{align}
 T_I^{\phantom{I}K} T_K^{\phantom{K}J} = \delta_I^{\phantom{I}J}. 
\label{T_property_sq}
\end{align}
Furthermore,
the following relations should hold 
since a hinge (or a triangle) whose arrows are all flipped 
is equivalent to a hinge (or a triangle) 
with indices in reverse order:
\begin{align}
 T_{I_1}^{\phantom{I_1}J_1} \ldots T_{I_k}^{\phantom{I_k}J_k} Y_{J_1 \ldots J_k} 
 &=  Y_{I_k \ldots I_1}  ,
\label{T_property_Y} 
\\
 C^{LMN}\,T_L^{\phantom{L}I} T_{M}^{\phantom{M}J} T_{N}^{\phantom{N}K} &=  C^{KJI} .
\label{T_property_C}
\end{align}

We define the Boltzmann weight $w(\gamma)$ of diagram $\gamma$ 
to be the product of $C^{IJK}$ and $Y_{I_1 \ldots I_k}$ 
followed by the summation over the indices on the edges:
\begin{align}
 w(\gamma)
 =\frac{1}{S(\gamma)}\,\sum_{\{I_e\}}
 \prod_{f:\,{\rm triangle}} C^{IJK}(f)\,
 \prod_{h:\,{\rm hinge}}Y_{I_1 \ldots I_k}(h)\,.
\end{align}
Here, $I_e$ are the indices on the edges, 
and $S(\gamma)$ is the symmetry factor of the diagram. 
The indices in $C^{IJK}$ and $Y_{I_1\ldots I_k}$ are contracted 
when the corresponding edges are identified 
(with $T_I^{\phantom{I}J}$ inserted appropriately if necessary).

The above diagrams with the prescribed Boltzmann weights 
can be generated as Feynman diagrams from the action%
\footnote{
Note that the 2-hinges (hinges with two edges) have been included as vertices.
This means that the set of the resulting diagrams 
contain the full set of triangular decompositions of two-dimensional surfaces. 
However, as we discuss in subsection \ref{tetrahedron}, 
the introduction of color structure excludes all such diagrams except for a tetrahedron, 
which is then interpreted as representing three-sphere $S^3$ 
obtained by gluing two tetrahedra face to face. 
} 
 
\begin{align}
 S[A,B] = \frac{1}{2} A_I B^I -\frac{\lambda}{6} C^{IJK} A_I A_J A_K
 - \sum_{k \geq 2}^{\infty} \frac{\mu_k}{2k} B^{I_1} \cdots B^{I_k} Y_{I_1 \ldots I_k} ,
\label{hm1_action}
\end{align}
where the dynamical variables $A_I$ and $B^I$ satisfy the relations 
\begin{align}
A_I= T_I^{\phantom{I}J} A_J, \qquad 
B^I= B^J T_J^{\phantom{J}I} .
\label{T_property_AB}
\end{align}
We have included the coupling constants $\lambda$ and $\mu_k$ $(k \geq 2)$ 
to count the numbers of triangles and $k$-hinges, respectively. 
In order to specify the directions of arrows, 
the index line will be a double line 
by setting index $I$ to be double index $I=(i,j)$.

It may be already clear, 
but we here explain how the action generates the diagrams we are concerned with. 
There are two kinds of interaction terms, 
one corresponding to triangles $C^{IJK}$ 
and the other to $k$-hinges $Y_{I_1 \ldots I_k}$ $(k\geq 2)$. 
The kinetic term $(1/2)A_I B^I$ yields a propagator 
that glues an edge of a triangle and that of a hinge. 
Note that 
two triangles cannot be glued to each other without an intermediate hinge, 
and two hinges cannot be glued to each other without an intermediate triangle.  
In order to handle the case where the tensor $T_I^{\phantom{I}J}$ needs to be inserted, 
we should multiply every leg of interaction terms  
by the factor $\delta_I^{\phantom{I}J} + T_I^{\phantom{I}J}$. 
However, this is equivalent to inserting the projector 
$\bigl(\delta_I^{\phantom{I}J} + T_I^{\phantom{I}J}\bigr)/2$ 
in every propagator, 
which in turn is equivalent to requiring 
that the dynamical variables be invariant 
under the action of $T_I^{\phantom{I}J}$.

In summary, our model is characterized by the data 
$(C^{IJK},Y_{I_1 \ldots I_k}, T_I^{\phantom{I}J})$ 
that satisfy the constraints 
\eqref{T_property_sq}--\eqref{T_property_C}.
In the next subsection 
we show that most of the constraints can be solved 
by considering semisimple associative algebras.

\subsection{Algebraic construction} 
\label{alg_construction} 

In this subsection, 
we give an algebraic construction of the model data 
$(C^{IJK},Y_{I_1 \ldots I_k}, T_I^{\phantom{I}J})$ 
(see \cite{Chung:1993xr} and also \cite{Fukuma:1993hy, Bachas:1992cr} 
for a related idea).

Let $\mathcal{R}$ be a real semisimple associative algebra.%
\footnote{
The following construction does not involve the operation of complex conjugation 
and thus can be readily generalized 
to associative algebras over the complex field.
} 
That is, $\mathcal{R}$ is a linear space over $\mathbb{R}$ 
with multiplication 
(denoted by $\times$) 
that satisfies the associativity, 
$(B_1\times B_2)\times B_3=B_1\times(B_2\times B_3)$. 
If one introduces a basis $\{E_I\}$ as $\mathcal{R} = \bigoplus_I \mathbb{R}\, E_I$, 
the multiplication is expressed in the form 
$E_I \times E_J = Y_{IJ}^{\phantom{IJ}K} E_K$, 
where the structure constants $Y_{IJ}^{\phantom{IJ}K}$ 
satisfy the relations 
$Y_{IJ}^{\phantom{IJ}L}Y_{LK}^{\phantom{LK}M}
 = Y_{IL}^{\phantom{IL}M}Y_{JK}^{\phantom{JK}L}$ 
due to associativity. 
The $k$-hinge tensor $Y_{I_1 \ldots I_k}$ can then be constructed 
from $Y_{IJ}^{\phantom{IJ}K}$ as
\begin{align}
 Y_{I_1 \ldots I_k} \equiv Y_{I_1 J_1}^{\phantom{I_1 J_1}J_k}
 Y_{I_2 J_2}^{\phantom{I_2 J_2}J_1} \ldots  Y_{I_k J_k}^{\phantom{I_k J_k}J_{k-1}} .
\label{Y_def}
\end{align}
It is easy to see that $Y_{I_1 \ldots I_k}$ are cyclically symmetric.%
\footnote{
The $k$-hinge tensor can also be expressed as 
$Y_{I_1 \ldots I_K}={\rm Tr}_{\,\mathcal{R}}\,
\bigl(E^{\rm reg}_{I_1}\cdots E^{\rm reg}_{I_k}\bigr)$, 
where $E^{\rm reg}_I$ are the  representation matrix of the basis $\{E_I\}$ 
in the regular representation of $\mathcal{R}$; 
$E^{\rm reg}_I=\bigl((E^{\rm reg}_I)^J_{\phantom{J}K}
=Y_{IK}^{\phantom{IK}J}\bigr)$ \cite{Fukuma:1993hy}.  
} 
The two-hinge tensor $Y_{IJ}$ is called the {\em metric} of $\mathcal{R}$ 
and will often be denoted by $G_{IJ}$; 
$G_{IJ}=Y_{IJ}=Y_{IK}^{\phantom{IK}L} Y_{JL}^{\phantom{JL}k}$.
It is known \cite{Fukuma:1993hy} that 
the associative algebra $\mathcal{R}$ is semisimple
(i.e., isomorphic to a direct sum of matrix rings) 
if and only if $G=(G_{IJ})$ has its inverse $G^{-1}\equiv(G^{IJ})$. 
The constraints \eqref{T_property_sq} and \eqref{T_property_Y} can be solved 
if there exists an involutory {\em anti}\,automorphism $T:\,\mathcal{R}\to\mathcal{R}$. 
In fact, the coefficients $T_I^{\phantom{I}J}$ in $T(E_I)=T_I^{\phantom{I}J}\,E_J$ 
satisfies \eqref{T_property_sq} when $T$ is involutory. 
Furthermore, when $T$ is an antiautomorphism: 
$T(E_J\times E_I)=T(E_I)\times T(E_J)$, 
we have the relations  
$T_I^{\phantom{I}K} T_J^{\phantom{J}L} Y_{KL}^{\phantom{KL}N}
=Y_{JI}^{\phantom{JI}M} T_M^{\phantom{M}N}$, 
which ensure \eqref{T_property_Y} to hold.

Such an  antiautomorphism can be naturally constructed 
when we set the index $I$ to be a double index $I=(i,j)$ $(i,j=1, \ldots ,N)$ 
in order to assign arrows to the edges of triangles and hinges. 
To see this, we let $\mathcal{R}$ take the form
\begin{align}
 \mathcal{R} &= \mathcal{A} \otimes \mathcal{\bar{A}} , 
\end{align}
where $\mathcal{A}$ and $\mathcal{\bar{A}}$ 
are linear spaces of the same dimension $N$. 
We fix an isomorphism from $\mathcal{A}$ to $\mathcal{\bar{A}}$ 
and denote it by $\sigma$.%
\footnote{
$\sigma$ can be taken arbitrarily 
because it can be absorbed into an automorphism of $\mathcal{A}$ or $\bar{\mathcal{A}}$.
} 
We assume that $\mathcal{A}$ is a semisimple associative algebra 
with multiplication $\times$.  
We introduce a multiplication (also denoted by $\times$) 
to $\mathcal{\bar{A}}$ 
such that $\sigma:\,\mathcal{A}\to\mathcal{\bar{A}}$ is 
an algebra {\em anti}\,automorphism, 
$\sigma(a\times b)=\sigma(b)\times \sigma(a)$ $(\forall a,\,b\in\mathcal{A})$.%
\footnote{
Such multiplication exists uniquely for a given $\sigma$ 
[see \eqref{ybar}].
} 
Then $\mathcal{R}=\mathcal{A}\otimes\mathcal{\bar{A}}$ 
naturally becomes an associative algebra 
as the tensor product of two associative algebras. 
The antiautomorphism $T:\,\mathcal{R}\to\mathcal{R}$ 
now can be defined 
such as to map an element $B=\sum b\otimes \bar{b}\in \mathcal{R}$ to
\begin{align}
 T(B)=\sum T(b\otimes \bar{b})\equiv \sum \sigma^{-1}(\bar{b})\otimes \sigma(b).
\end{align}
One can easily show that $T$ is certainly an antiautomorphism. 
We thus find that the constraints \eqref{T_property_sq} and \eqref{T_property_Y} 
can be solved by giving an associative algebra $\mathcal{A}$.

Note that $\mathcal{R}$ can also be thought of 
as the set of algebra endomorphisms of $\mathcal{A}$ 
by regarding $\mathcal{\bar{A}}$ as the dual linear space of $\mathcal{A}$\,: 
\begin{align}
 \mathcal{R} &= \mathcal{A} \otimes \mathcal{\bar{A}} = {\rm End}\,\mathcal{A}. 
\label{endomorphisms}
\end{align}
One then can also define an antiautomorphism $\bar{T}$ for the dual of $\mathcal{R}$, 
$\bar{\mathcal{R}}\equiv \bar{\mathcal{A}}\otimes \mathcal{A}$, 
such that the following relation holds:
\begin{align}
 \langle \,\bar{T}(A),\,T(B) \rangle
 =\langle A,\,B \rangle
 \quad (\forall A\in\bar{\mathcal{R}},\,\forall B\in\mathcal{R}),
\end{align}
where $\langle~,~\rangle$ is the paring between $\bar{\mathcal{R}}$ 
(the dual of $\mathcal{R}$) and $\mathcal{R}$.

We rephrase the above construction in terms of the bases $\{e_i\}$ and $\{\bar{e}^i\}$ 
of $\mathcal{A}$ and $\mathcal{\bar{A}}$\,: 
\begin{align}
 \mathcal{A} &= \bigoplus_{i=1}^N \mathbb{R}\, e_i,   
 \qquad \mathcal{\bar{A}} = \bigoplus_{i=1}^N \mathbb{R}\, \bar{e}^i . 
\end{align}
We first represent the isomorphism $\sigma$ as
\begin{align}
 \sigma(e_i) = \sigma_{ij}\,\bar{e}^j , \qquad
 \sigma^{-1}(\bar{e}^i) = \sigma^{ij}\,e_j ,
\label{def_sigma}
\end{align}
and write the structure constants of the multiplication on $\mathcal{A}$ as 
$e_i \times e_j = y_{ij}^{\phantom{ij}k}e_k$. 
Then those of $\mathcal{\bar{A}}$ 
(appearing in $\bar{e}^i \times \bar{e}^j = \bar{y}^{ij}_{\phantom{ij}k} \bar{e}^k$) 
are determined from the requirement of antihomomorphism, 
$\sigma(e_i \times e_j)= \sigma(e_j) \times \sigma(e_i)$, 
to be
\begin{align}
 \bar{y}^{ij}_{\phantom{ij}k} = \sigma^{il} \sigma^{jm} y_{ml}^{\phantom{ml}n} \sigma_{nk}.
\label{ybar}
\end{align}
If we take the basis of $\mathcal{R}$ to be $E_I=E_i^{\phantom{i}j}=e_i \otimes \bar{e}^j$, 
then the structure constants $Y_{I_1 I_2}^{\phantom{I_1I_2}I_3}$ are given by
\begin{align}
 Y_{I_1 I_2}^{\phantom{I_1I_2}I_3} =Y_{i_1}^{\phantom{i_1}j_1}\!~_{i_2  }^{\phantom{i_2} j_2}\!~^{i_3}_{\phantom{i_3}j_3}
 =y_{i_1i_2}^{\phantom{i_1i_2}i_3}\,\bar{y}^{j_1j_2}_{\phantom{j_1j_2}j_3},
\label{Yyy}
\end{align}
from which the $k$-hinge tensor is given by
\begin{align}
 Y_{I_1 \ldots I_k}=Y_{i_1}^{\phantom{i_1}j_1}\cdots\!~_{i_k}^{\phantom{i_k}j_k}
=y_{i_1 \ldots i_k}\,\bar{y}^{j_1 \ldots j_k}
\end{align}
with
\begin{align}
 y_{i_1 \ldots i_k} &\equiv y_{i_1 j_1}^{\phantom{i_1 i_1}j_k} 
 y_{i_2 j_2}^{\phantom{i_2 j_2}j_1} \ldots y_{i_k j_k}^{\phantom{i_k j_k}j_{k-1}} ,
\label{subscript_y}
\\
 \bar{y}^{i_1 \ldots i_k} &\equiv 
 \bar{y}^{i_1 j_1}_{\phantom{i_1 i_1}j_k} 
 \bar{y}^{i_2 j_2}_{\phantom{i_2 j_2}j_1} \ldots 
 \bar{y}^{i_k j_k}_{\phantom{i_k j_k}j_{k-1}}
 =\sigma^{i_1 j_1}\cdots \sigma^{i_k j_k} y_{j_k\ldots j_1}. 
\end{align}
In particular, the metric of $\mathcal{R}$ takes the form  
\begin{align}
 G_{I_1 I_2}=G_{i_1}{}^{j_1}{}_{i_2}{}^{j_2}
 =g_{i_1 i_2}\,\bar{g}^{j_1 j_2},
\end{align}
where $g_{i_1 i_2}$ and $\bar{g}^{j_1 j_2}$ are the metrics 
of $\mathcal{A}$ and $\mathcal{\bar{A}}$, respectively; 
$g_{i_1 i_2}\equiv y_{i_1 k}^{\phantom{i_1 k}\ell}\,y_{i_2 \ell}^{\phantom{i_1 \ell}k}$, 
$\bar{g}^{j_1 j_2}=\bar{y}^{j_1 k}_{\phantom{j_1 k}\ell}\,
\bar{y}^{j_2 \ell}_{\phantom{i_1 \ell}k}$.%
\footnote{
Note that  the cyclically symmetric tensor $y_{i_1 \ldots i_k}$ can also be written as 
\begin{align}
 y_{i_1 i_2 i_3}=y_{i_1 i_2}^{ \phantom{i_1 i_2} j_3}g_{j_3 i_3}, 
 \quad
 y_{i_1 \ldots i_k} = y_{i_1j_1 l_1}\,g^{j_1 l_2} y_{i_2j_2 l_2}\,g^{j_2 l_3}
 \ldots y_{i_k j_k l_k}\,g^{j_k l_1} .
\nn
\end{align}
} 
We easily see that $\mathcal{R}$ is semisimple if $\mathcal{A}$ is, 
because $G_{I_1 I_2}$ has its inverse 
when $g_{i_1 i_2}$ does (and so does $\bar{g}^{j_1 j_2}$).

The antiautomorphism $T_I^{\phantom{I}J} $ is now expressed as 
$T(e_i\otimes \bar{e}^j)\equiv \sigma^{-1}(\bar{e}^j) 
\otimes \sigma(e_i)=\sigma^{jk}\sigma_{il} \, e_k\otimes \bar{e}^l $, 
that is, 
\begin{align}
 T_{I_1}^{\phantom{I_1}I_2}
 =T_{i_1}^{\phantom{i_1}j_1}{}^{i_2}_{\phantom{i_2}j_2}
 =\sigma_{i_1 j_2}\,\sigma^{j_1 i_2}.
\end{align}
For the dual algebra $\bar{\mathcal{R}}=\bar{\mathcal{A}}\otimes\mathcal{A}$, 
regarding $\{\bar{e}^i\}$ as the dual basis of $\{e_i\}$, 
we set a basis of $\bar{\mathcal{R}}$ to be 
$\bar{E}^I= \bar{E}^i_{\phantom{i}j} = \bar{e}^i \otimes e_j$, 
which leads to the pairing 
$\langle\bar{E}^{I_1},\,E_{I_2}\rangle
=\delta^{i_1}_{i_2}\,\delta^{j_2}_{j_1}$. 
Then the antiautomorphism $\bar{T}$ on $\bar{\mathcal{R}}$ 
is expressed as 
$\bar{T}(\bar{E}^I) \equiv \bar{E^J} (T^{-1})_{J}^{\phantom{J}I}
=\bar{E^J} T_{J}^{\phantom{J}I}$.

The dynamical variables $A_I$ and $B^I$ in \eqref{hm1_action} 
can be regarded as elements of $\bar{\mathcal{R}}$ and $\mathcal{R}$, respectively:
\begin{align}
 A= A_I \bar{E}^I =A_{i}^{\phantom{i}j}\, \bar{e}^i\otimes e_j \in \bar{\mathcal{R}}, \qquad
 B=B^I E_I = B^{i}_{\phantom{i}j}\, e_i\otimes \bar{e}^j  \in  \mathcal{R}.
\end{align}
The condition \eqref{T_property_AB} is then expressed as $\bar{T}(A)=A $, $T(B) = B$.  
With these double indices, the action \eqref{hm1_action} is written as
\begin{align}
 S = \frac{1}{2}A_i^{\phantom{i}j}B^i_{\phantom{i}j} 
 - \frac{\lambda}{6} C^{i\phantom{j}k\phantom{l}m\phantom{n}}_{\phantom{i}j\phantom{k}l\phantom{m}n} 
 A_i^{\phantom{i}j}A_k^{\phantom{k}l}A_m^{\phantom{m}n} 
 -\sum_{k \geq 2} \frac{\mu_k}{2k} B^{i_1}_{\phantom{i_1}j_1} \ldots 
 B^{i_k}_{\phantom{i_k}j_k}y_{i_1 \ldots i_k} \bar{y}^{j_1 \ldots j_k},
\end{align}
where the tensor $C^{i\phantom{j}k\phantom{l}m\phantom{n}}_{\phantom{i}j\phantom{k}l\phantom{m}n}$ is
arbitrary as long as it satisfies the condition \eqref{T_property_C}.
It is often convenient to use 
$A_{ij} \equiv \sigma_{jk} A_i^{\phantom{i}k}$, 
$B^{ij} \equiv B^i_{\phantom{i} k}\sigma^{kj}$, 
and
$C^{ijklmn} \equiv 
C^{i\phantom{j^\prime} k \phantom{l^\prime} m 
\phantom{n^\prime}}_{\phantom{i} j^\prime \phantom{k} l^\prime \phantom{m} n^\prime} 
\sigma^{j^\prime j} \sigma^{l^\prime l} \sigma^{n^\prime n}$. 
Then the conditions \eqref{T_property_C} and \eqref{T_property_AB} 
can be rewritten to the form where $\sigma$ does not appear:  
\begin{align}
&C^{ijklmn} = C^{nmlkji} ,  
\label{property_C_supsc} \\
&A_{ij} = A_{ji}, \quad B^{ij} = B^{ji} .  
\label{sym_AandB}
\end{align}
The action then becomes 
\begin{align}
 S = \frac{1}{2} A_{ij}B^{ij} - \frac{\lambda}{6} C^{ijklmn}A_{ij}A_{kl}A_{mn} 
 - \sum_{k \geq 2} \frac{\mu_k}{2k} B^{i_1 j_1} \cdots B^{i_k j_k}
 y_{i_1 \ldots i_k} y_{j_k \ldots j_1} 
\label{hm2_action}.
\end{align}
Thus, a set of models can be defined by giving semisimple associative algebras $\mathcal{A}$ 
and the tensors $C^{ijklmn}$ satisfying \eqref{property_C_supsc}. 
The isomorphism $\sigma:\,\mathcal{A}\to\bar{\mathcal{A}}$ can be taken arbitrarily   
and is regarded as a sort of gauge freedom 
in choosing the basis of $\mathcal{A}$ or $\bar{\mathcal{A}}$.

\subsection{The Feynman rules}
\label{feynman_rule}

As stated in the last subsection, 
the tensor $C^{ijklmn}$ in \eqref{hm2_action} can be chosen arbitrarily 
as long as it satisfies the condition \eqref{property_C_supsc}. 
In this paper, we set it to be 
\begin{align}
 C^{ijklmn} = g^{jk}g^{lm}g^{ni} ,
\label{C_def}
\end{align}
which one can easily show to satisfy \eqref{property_C_supsc}.%
\footnote{
This choice \eqref{C_def} will be slightly modified 
when we introduce a color structure to our models. 
Other choices will be studied in our future paper \cite{fsu2}. 
} 
Then the Feynman rules of the action \eqref{hm2_action} become 
\begin{align}
 {\rm propagator}&:\ \begin{array}{l} \includegraphics[height = 1.2cm]{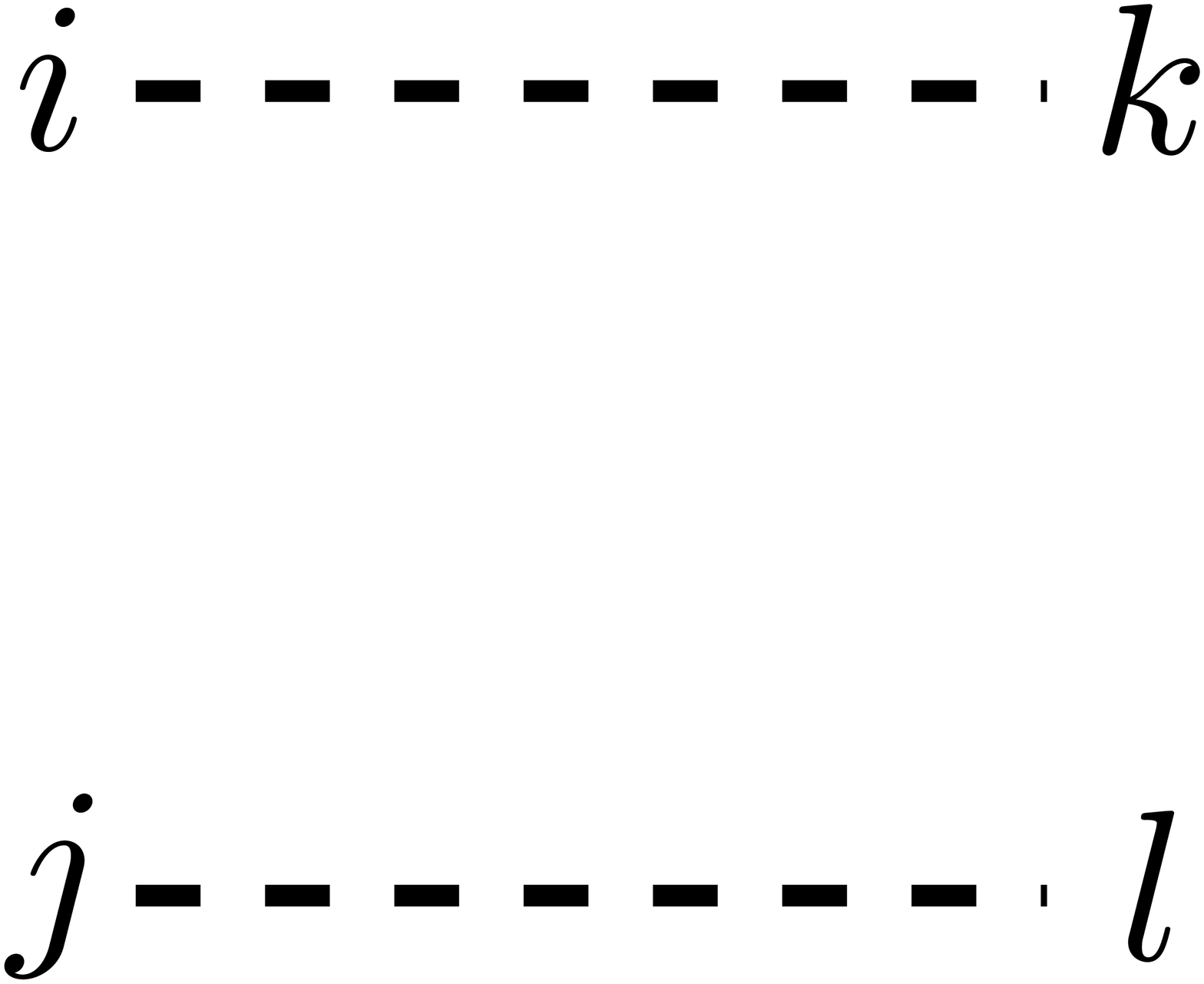} 
 \end{array} 
 + \begin{array}{l} \includegraphics[height = 1.2cm]{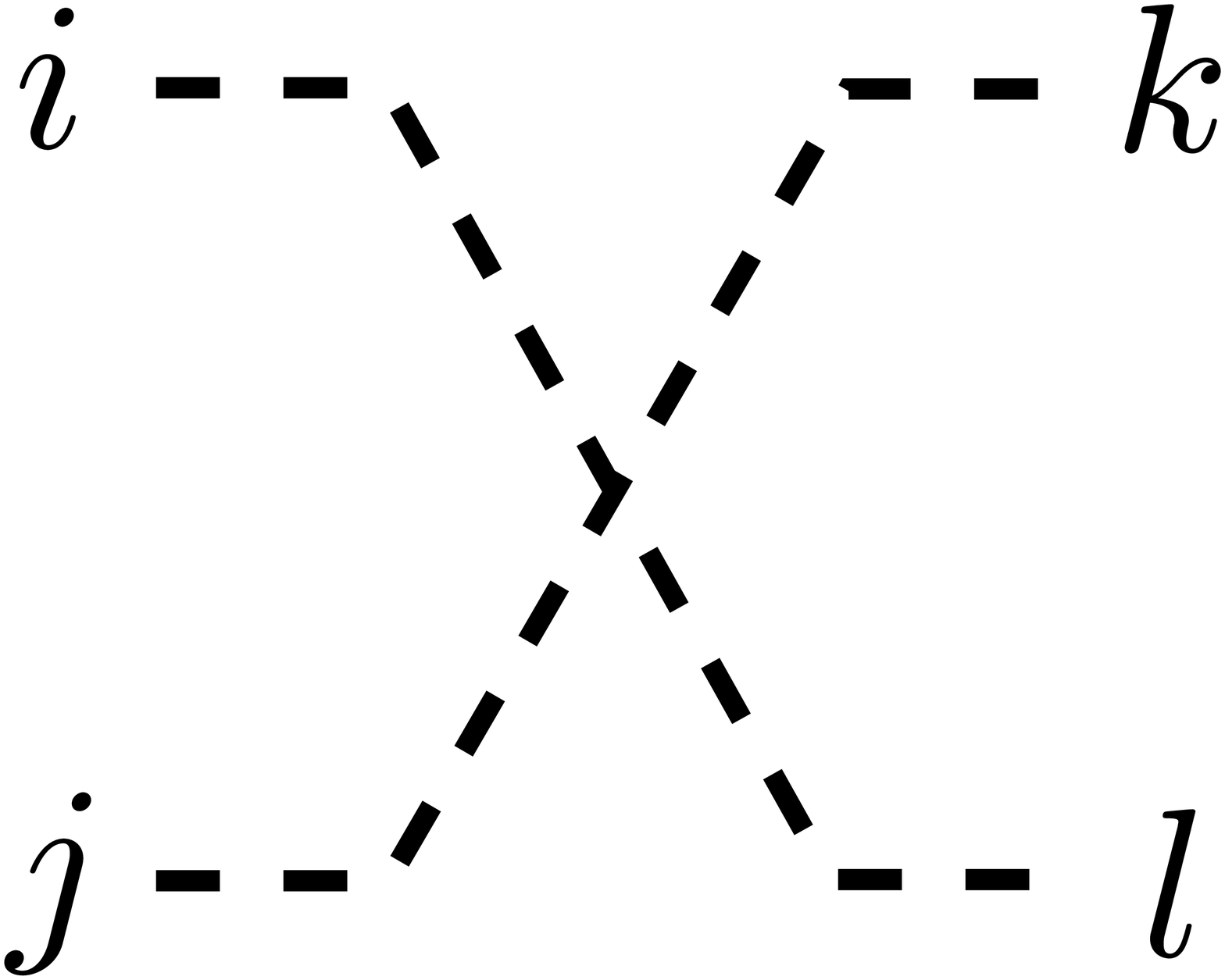}
 \end{array}  
 \sim \langle A_{ij}B^{kl} \rangle = \delta_i^{\phantom{i}k} \delta_j^{\phantom{j}l}
 + \delta_i^{\phantom{i}l} \delta_j^{\phantom{j}k} ,
\label{hm2_propagator} 
\\
 {\rm triangle}&:\ \begin{array}{l} \includegraphics[width = 2.0cm]{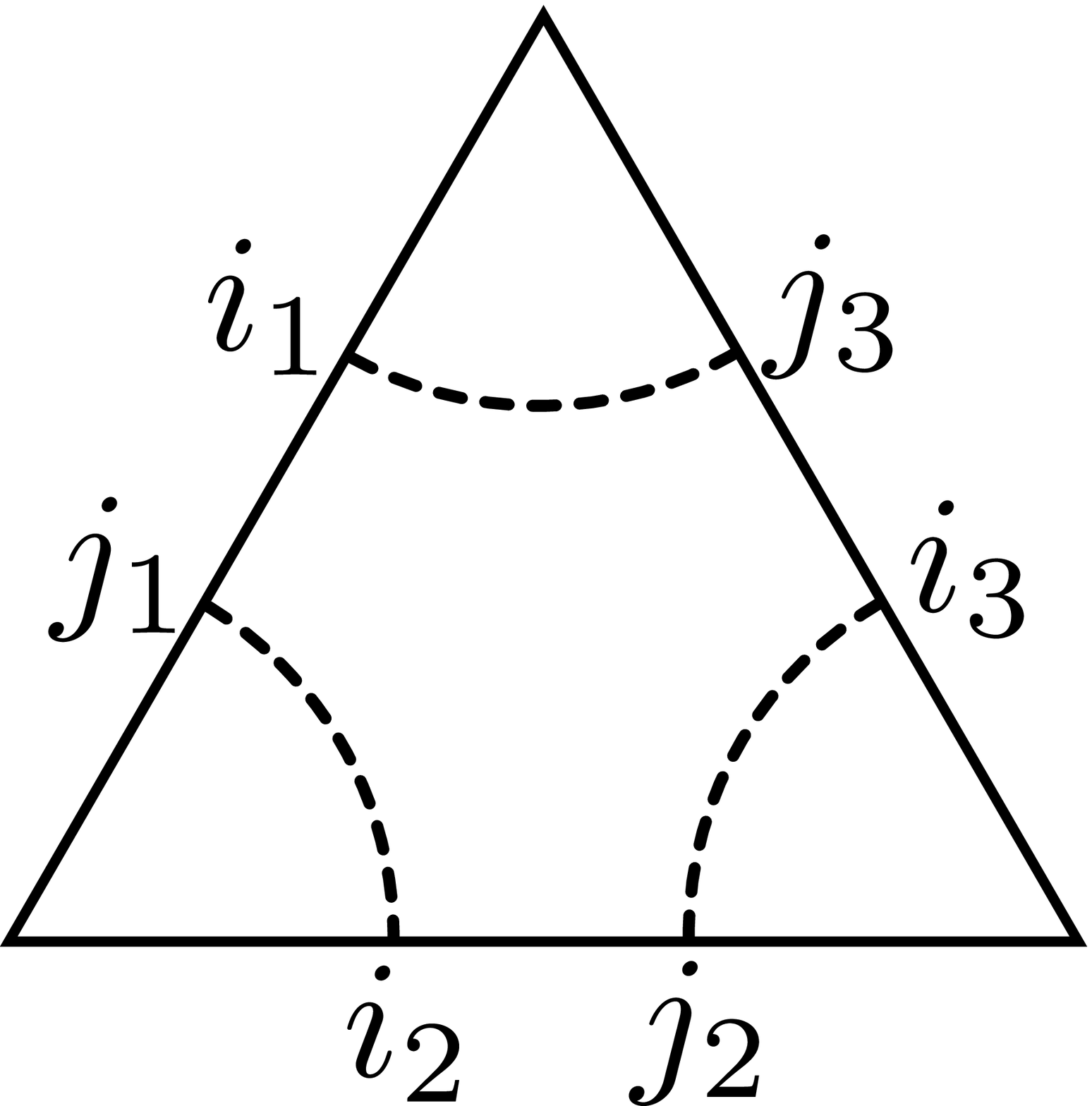}
 \end{array}
 \sim \lambda \, g^{j_1i_2}\,g^{j_2i_3}\,g^{j_3i_1} ,
\label{hm2_triangle} 
\\
 \mbox{$k$-hinge}&:\ \begin{array}{l} \includegraphics[width = 2.0cm]{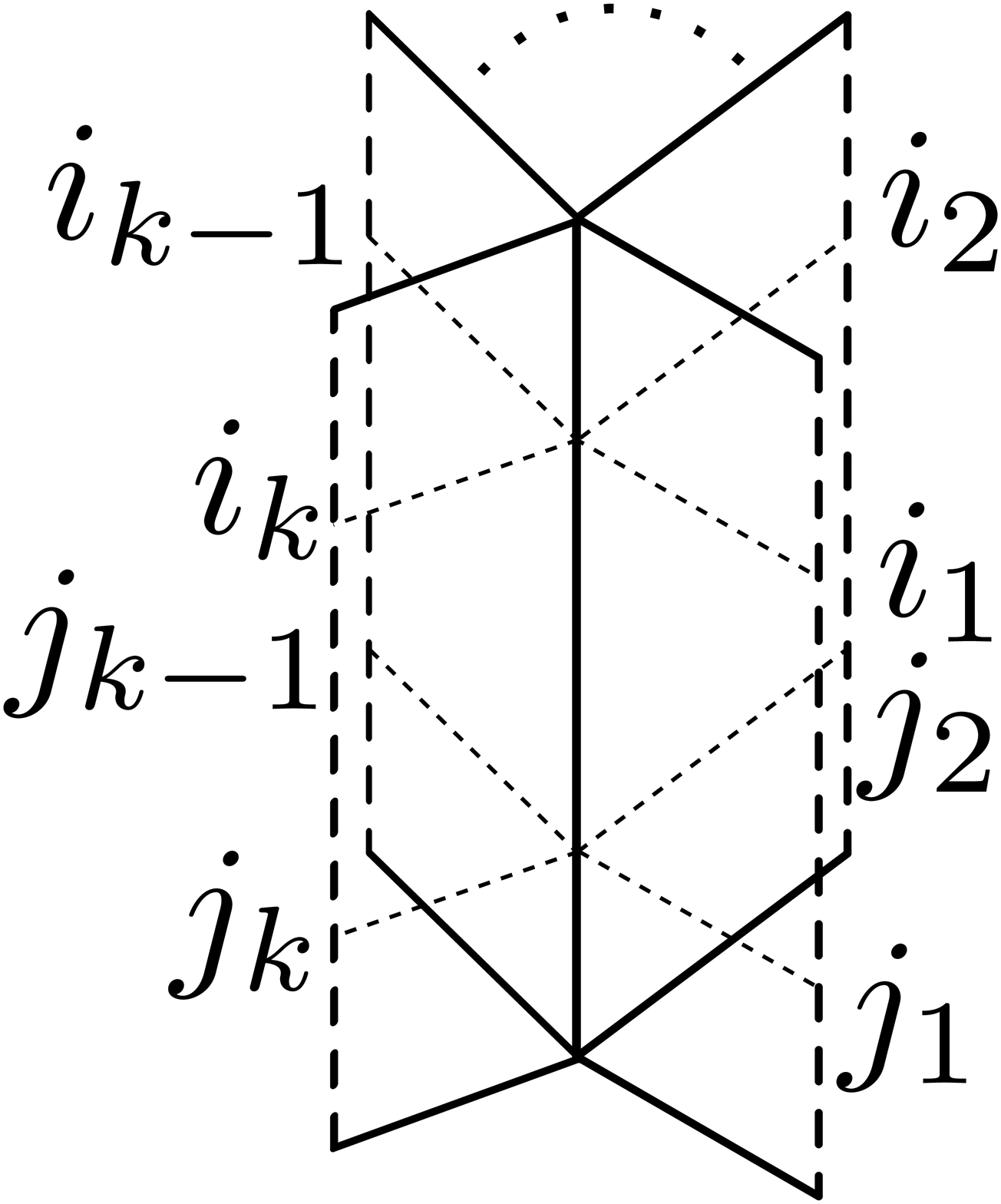}
 \end{array} \sim \mu_k \, y_{i_1 \ldots i_k} \,  y_{j_k \ldots j_1}  .
\label{hm2_hinge}
\end{align}
Recall that the arrows are now expressed with double lines, 
and thus, 
when the first (or second) term of the propagator \eqref{hm2_propagator} is used 
the edges are glued in the same (or opposite) direction.

The free energy of this model takes the form
\begin{align}
 \log Z = \sum_{\gamma} \frac{1}{S(\gamma)} \, \lambda^{s_2(\gamma)} 
 \Bigl( \prod_{k \geq 2} \mu_k^{s_1^k(\gamma)} \Bigr) \, \mathcal{F}(\gamma) .
\label{free_energy_general}
\end{align}
Here, the sum $\sum_\gamma$ is taken over all possible connected Feynman diagrams $\{\gamma\}$, 
and $S(\gamma)$ is the symmetry factor of diagram $\gamma$. 
$s_2(\gamma)$ is the number of triangles, 
and $s_1^k(\gamma)$ the number of $k$-hinges.
$\mathcal{F}(\gamma)$ denotes the product of $y_{i_1 \ldots i_k}$ and $g^{ij}$ 
with the indices contracted according to a given Wick contraction, 
and we call $\mathcal{F}(\gamma)$ 
the {\em index function of diagram} $\gamma$. 
We here regard two diagrams as being the same  
if the indices are contracted in the same manner. 
The numerical coefficients in the action \eqref{hm2_action} are chosen 
such that independent diagrams 
give only the symmetry factors to the free energy, 
by taking into account the symmetry (rotation and flip) of triangles and hinges.  
We stress that the free energy will have a different form from \eqref{free_energy_general} 
if we make a different choice for $C^{ijklmn}$ other than \eqref{C_def}.

\begin{figure}[htbp]
\begin{center}
\includegraphics[height = 4cm]{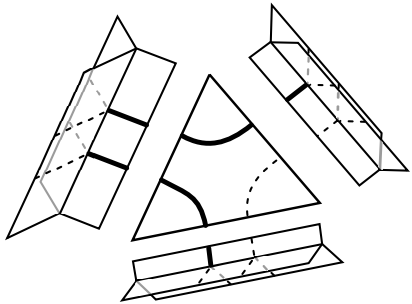}
\begin{quote}
\vspace{-5mm}
\caption{
A part of index networks. 
Two index lines (depicted in bold lines) come out of the same edge of the left hinge 
and enter the right adjacent triangle. 
The upper index line then leaves the triangle and enters the upper hinge, 
while the lower index line enters the lower hinge. 
}
\label{fig:part_of_index_network}
\end{quote}
\end{center}
\vspace{-6ex}
\end{figure}

\begin{figure}[htbp]
\begin{center}
\includegraphics[height = 4cm]{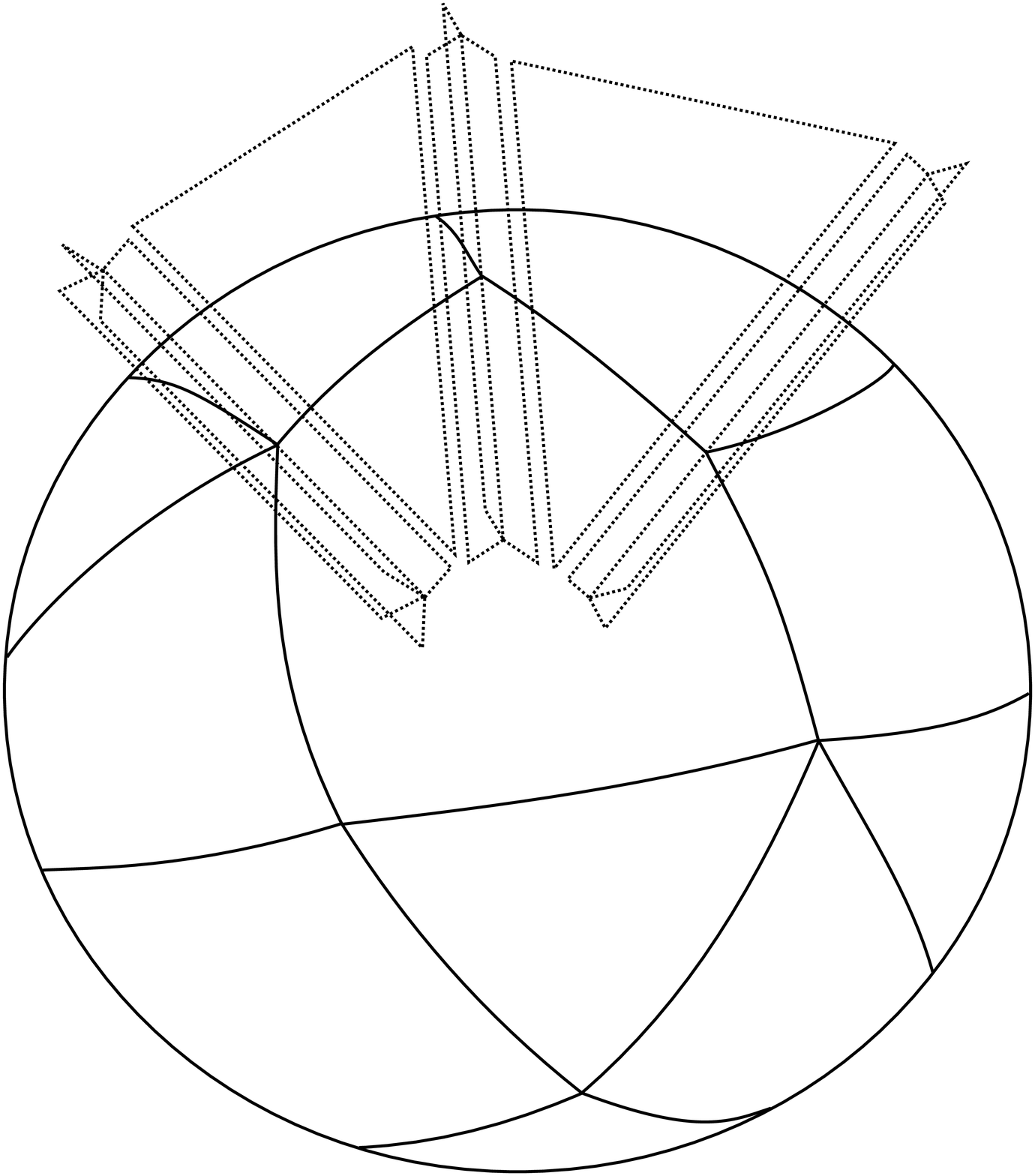}
\begin{quote}
\vspace{-5mm}
\caption{
Index network around a vertex. 
It represents a polygonal decomposition of a closed surface 
(not necessarily a sphere) around a vertex. 
A $k$-valent junction in the index network corresponds to a $k$-hinge in the original diagram,
where $k$ index lines meet. 
A segment connecting two junctions in the index network 
corresponds to an intermediate triangle between two hinges. 
}
\label{fig:index_network}
\end{quote}
\end{center}
\vspace{-6ex}
\end{figure}

We now show that the index function $\mathcal{F}(\gamma)$ can be expressed 
as the product of the contributions from {\em two-dimensional}\, surfaces, 
each surface enclosing a vertex of  diagram $\gamma$. 
To see this, 
we first note from the Feynman  rules \eqref{hm2_propagator}--\eqref{hm2_hinge} 
that even a connected Feynman diagram 
generally gives disconnected networks of index lines. 
This is because each hinge has a pair of junction points as for index lines 
and two index lines out of the same edge of a hinge can enter two different hinges 
after passing through an adjacent triangle (see Fig.~\ref{fig:part_of_index_network}). 
We further note that the index lines on two different hinges 
can be connected (through an intermediate triangle) 
if and only if the hinges share the same vertex of $\gamma$. 
This means that the {\em connected}\, index networks 
have a one-to-one correspondence to the vertices of $\gamma$. 
We thus find that the index function $\mathcal{F}(\gamma)$ of diagram $\gamma$ 
is the product of the contributions from connected index networks 
(each assigned to a vertex of $\gamma$) 
and has the form
\begin{align}
 \mathcal{F}(\gamma) = \prod_{v:\,\text{vertex of $\gamma$}} \zeta(v) .
\label{index_fcn_prod}
\end{align}
We also call $\zeta(v)$ the index function 
(more precisely, the {\em index function of vertex} $v$). 
Note that every connected index network 
takes the form of a polygonal decomposition of a closed surface 
(not necessarily a sphere and may include monogons or digons), 
where a $k$-valent junction (or $k$-junction) corresponds to a $k$-hinge 
where $k$ index lines meet 
(see Fig.~\ref{fig:index_network}).%
\footnote{
In order to avoid possible confusions 
between the terms for Feynman diagrams and those for index networks, 
we call the vertices and edges in the index network 
the junctions and segments, respectively. 
} 

We here make a few comments. 
The first comment is on the uniqueness 
in interpreting an index network 
as a polygonal decomposition of a closed surface. 
In fact, if one regards an index network 
simply as a wire frame (i.e., as a collection of segments), 
then it is not a unique procedure 
to assign polygonal faces in the frame 
such that the resulting configuration forms a closed surface.%
\footnote{
For example, 
there arises such ambiguity 
if a diagram includes a triangle shared by more than three tetrahedra, 
as in $n$-simplex ($n\geq 5$) 
which can be constructed 
from triangles and multiple hinges.
\label{fn:higher-dimensional-simplex}
} 
However, our index network is not simply a wire frame, 
and has the information on how the indices are contracted. 
We thus can uniquely assign faces to the holes of the index network 
by carefully following the contraction of indices. 
We will see in section \ref{hm2mr} 
that the assignment is straightforward 
when models are given by matrix rings 
as the defining associative algebras.

The second comment is on the manifoldness of a diagram $\gamma$. 
Since there is a two-dimensional surface around each vertex of $\gamma$,  
we can say that there is a three-dimensional cone at each vertex, 
the base and apex of a cone 
being the connected index network around a vertex 
and the vertex itself, respectively.  
For example, if an index network has the topology of two-sphere $S^2$, 
then the corresponding cone is a 3-dimensional ball $B^3$.  
These cones characterize the neighborhoods of the vertices of the diagram $\gamma$.%
\footnote{
Some diagram (as the one in footnote \ref{fn:higher-dimensional-simplex}) 
may be better regarded as being a higher dimensional object. 
In this case, the above three-dimensional cone 
will be treated as a part of the neighborhood of a vertex 
in the higher dimensional object.
} 
Note that $\gamma$ represents a three-dimensional (combinatorial) manifold 
if $\gamma$ gives a tetrahedral decomposition 
and the neighborhood of every vertex is homeomorphic to $B^3$. 
In section \ref{hm2mr}, 
by taking $\mathcal{A}$ to be a matrix ring 
and introducing a color structure to the models, 
we show that the set of possible Feynman diagrams can be drastically reduced 
such that only (and all of the) manifolds are generated.

\subsection{Evaluation of diagrams}
\label{evaluation}

The index function $\mathcal{F}(\gamma)$ can be easily evaluated
by deforming each connected index network  
with the use of the associativity of 
$y_{ij}^{\phantom{ij}k}=y_{ijl}\,g^{lk}$ \cite{Fukuma:1993hy}:
\begin{align}
 \begin{array}{l} \includegraphics[width = 2cm]{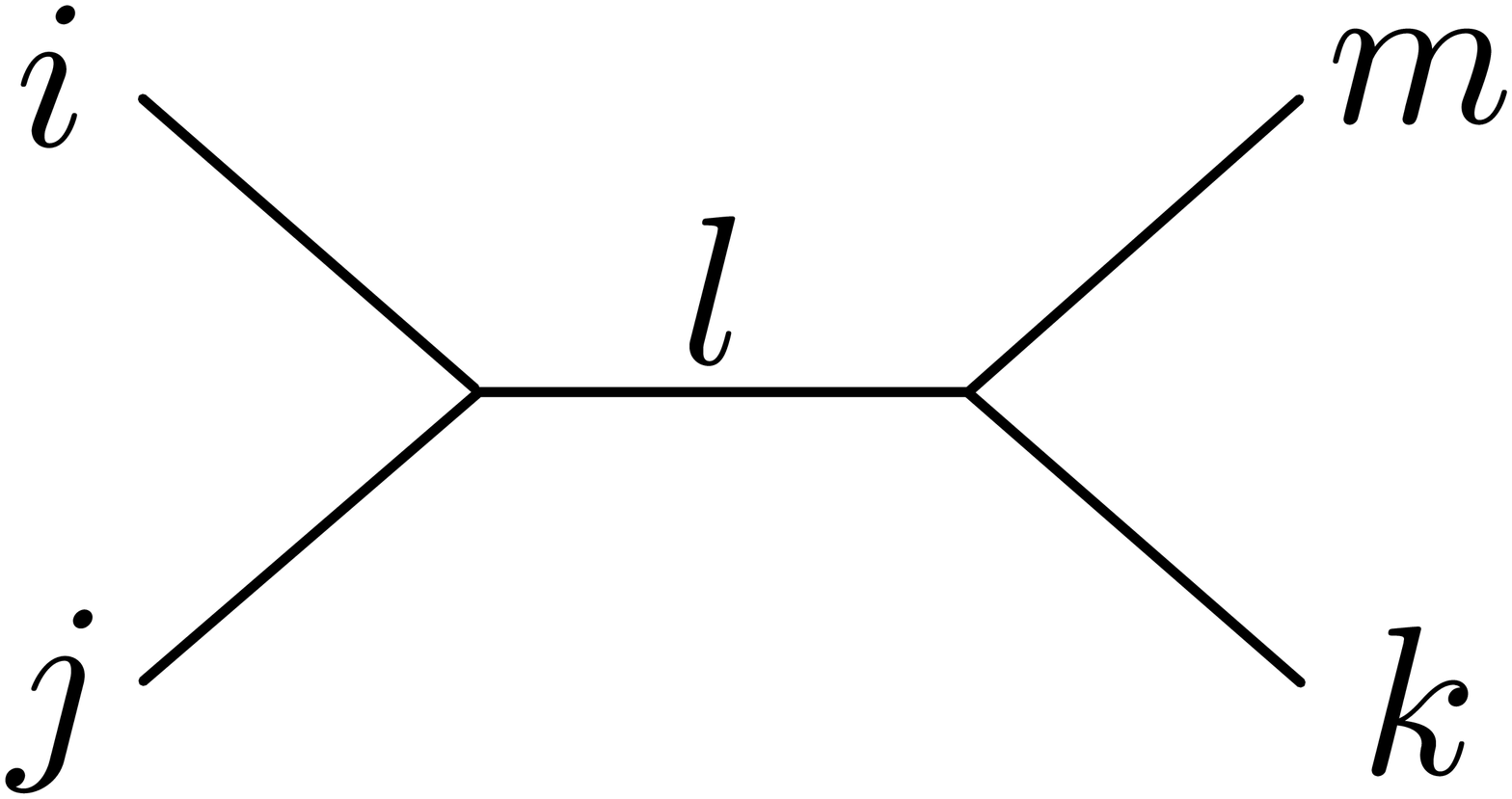} \end{array} 
 = y_{ij}^{\phantom{ij}l}y_{lkm}
 = y_{ilm} y_{jk}^{\phantom{jk}l}
 = \begin{array}{l} \includegraphics[height = 2cm]{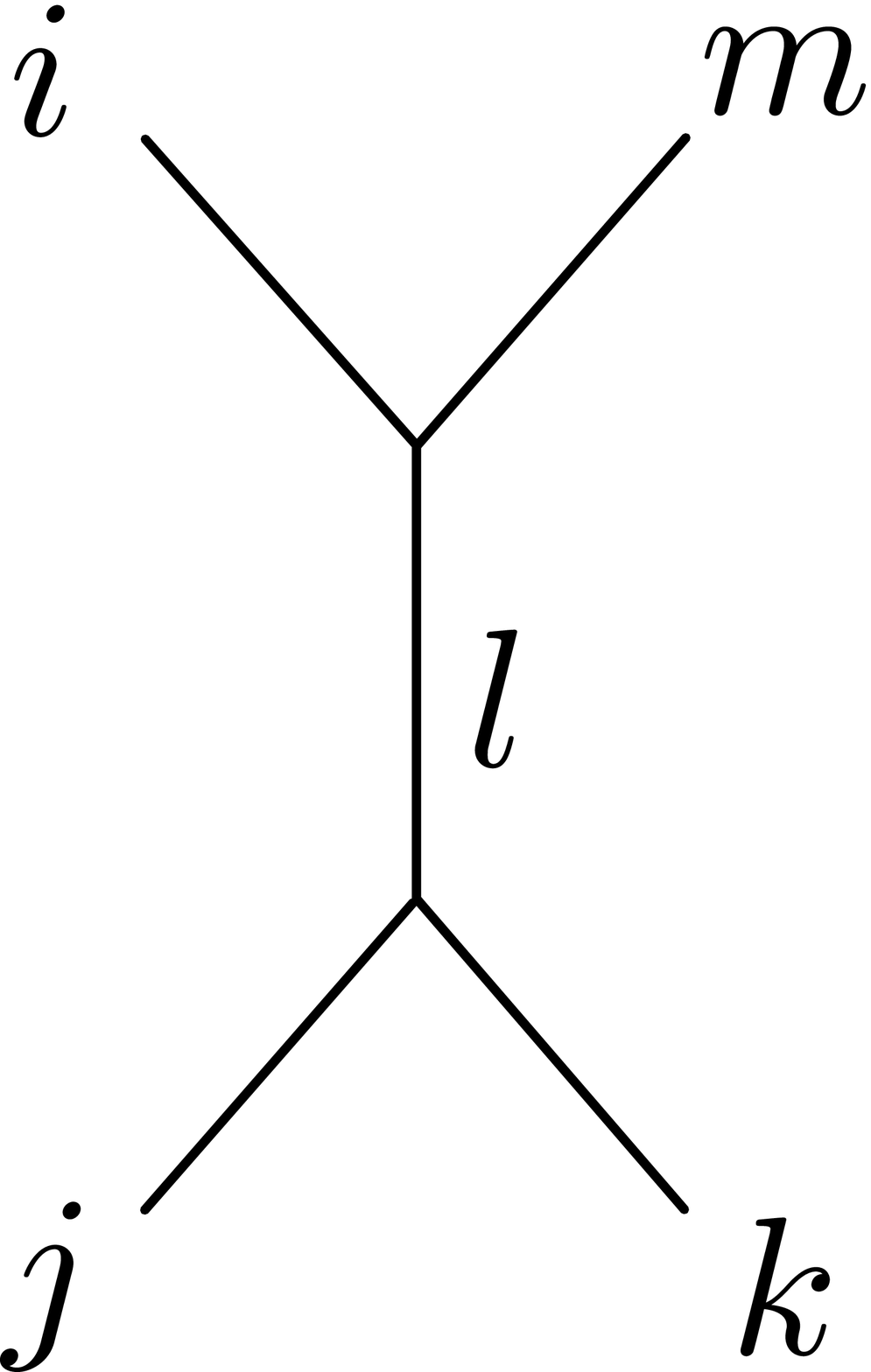}\end{array} .
\label{2-2}
\end{align}
In deformation 
there may appear two kinds of index loops: 
\begin{align}
 \begin{array}{l} \includegraphics[width = 4cm]{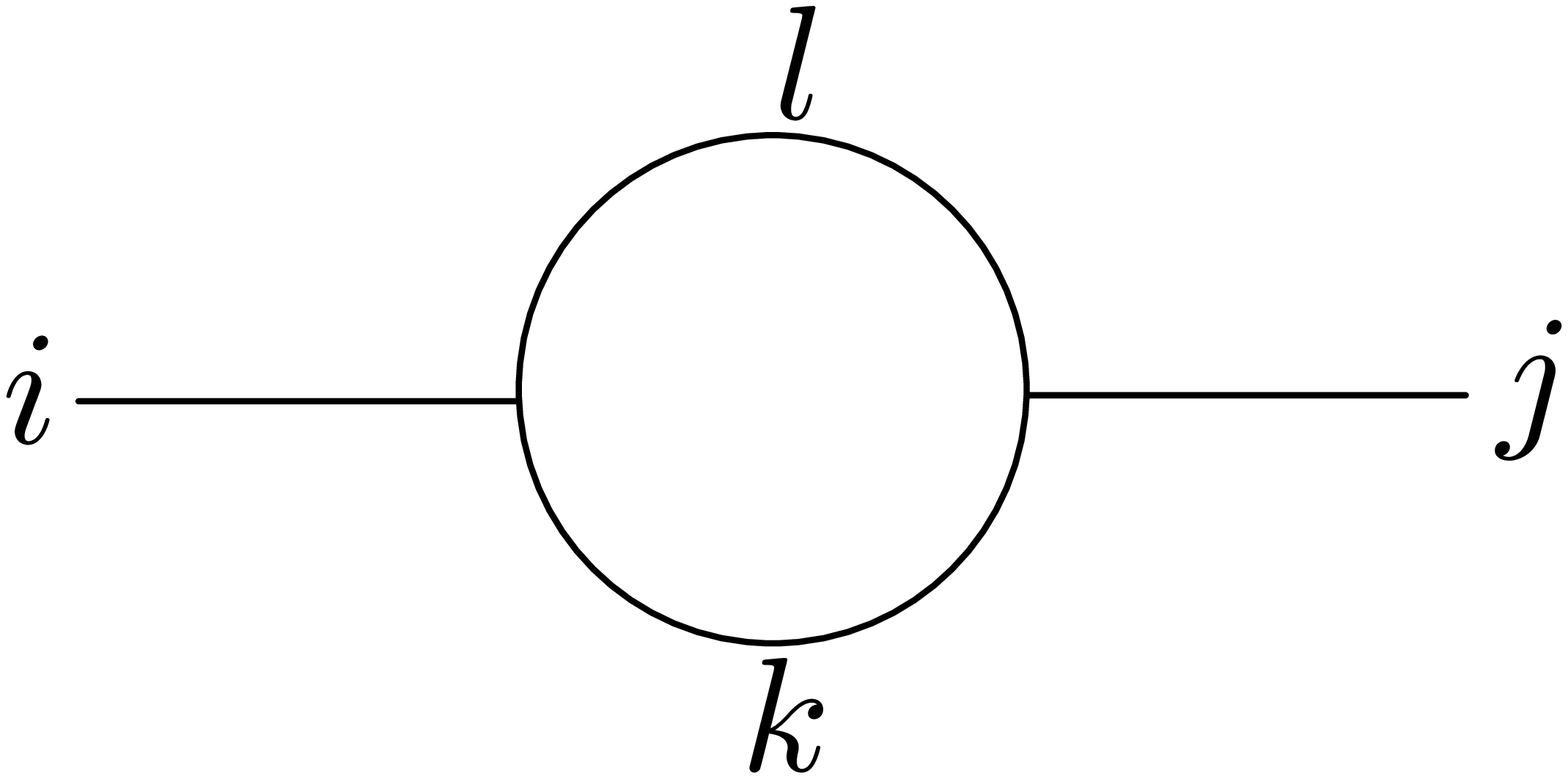} \end{array}
 &= y_{ik}^{\phantom{ik}l}\,y_{jl}^{\phantom{jl}k}
 ~~\Bigl(\,
 = g_{ij} 
 = \begin{array}{l} \includegraphics[width = 1.75cm]{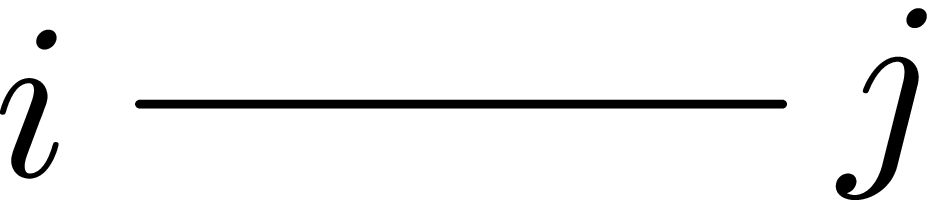} \end{array}
 \Bigr), 
\label{2d-bubble}
\\
 \begin{array}{l} \includegraphics[width = 4cm]{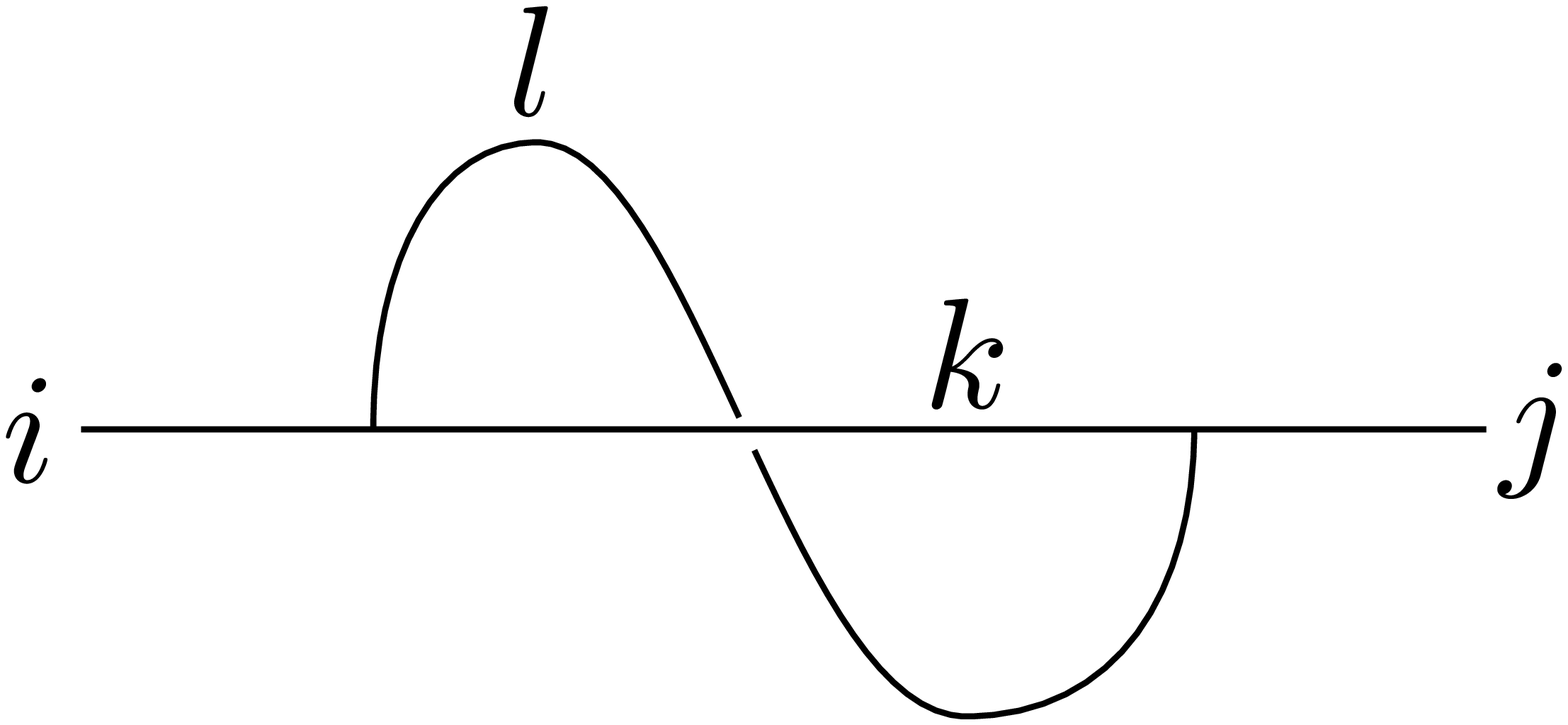} \end{array}
 &= y_{ik}^{\phantom{ik}l}\,y_{lj}^{\phantom{lj}k}
 ~~\Bigl(\,
 \equiv p_{ij} 
 \Bigr).
\label{proj_p}
\end{align}
The former index loop diagram can be replaced by a single solid line, 
while the loop in the latter index diagram cannot be removed. 
Actually $p_{ij}$ (or more precisely, $p_i^j\equiv p_{ik} g^{kj}$) 
is a projector to the center of algebra $\mathcal{A}$, 
$Z(\mathcal{A})$,
as can be checked easily \cite{Fukuma:1993hy}. 
If a given connected index network does not produce a projector $p_{ij}$ 
in the process of deformation, 
the index network can always be deformed to a single circle 
after repeatedly using \eqref{2-2} and \eqref{2d-bubble} 
and gives the value $g_{ij}g^{ij}=N$, 
where $N$ is the dimension of $\mathcal{A}$. 
On the other hand, 
if a given connected index network admits the appearance of a projector $p_{ij}$, 
the value of the index network is generally less than $N$.%
\footnote{
For example, $p_{ij} g^{ij}=p^i_i$ gives the linear dimension 
of $Z(\mathcal{A})$. 
} 

Note that the two deformations \eqref{2-2} and \eqref{2d-bubble} 
are actually the local moves of two-dimensional surfaces.%
\footnote{
Namely, any two index networks can be obtained from each other 
by a repetitive use of \eqref{2-2} and \eqref{2d-bubble}  
if and only if the two index networks represent  
two-dimensional surfaces of the same topology \cite{Fukuma:1993hy}.
} 
Therefore, the index function $\zeta(v)$ of vertex $v$ gives 
a two-dimensional topological invariant defined by the associative algebra $\mathcal{A}$ 
\cite{Fukuma:1993hy}, 
and thus has the form $\zeta(v)=\mathcal{I}_{g(v)}$, 
where $g(v)$ is the genus of the network.%
\footnote{
We already know some of the general results, 
$\mathcal{I}_{g=0}=\text{dim}\mathcal{A}=N$, 
$\mathcal{I}_{g=1}=\text{dim}Z(\mathcal{A})$. 
} 
Thus the index function of diagram $\gamma$ is expressed as 
\begin{align}
 \mathcal{F}(\gamma) = \prod_{v:\,\text{vertex}} \zeta(v)
 = \prod_{v:\,\text{vertex}} \mathcal{I}_{g(v)} .
\label{index_fcn_prod_topological}
\end{align}

\subsection{Examples} 
\label{diagram_example}

In this subsection we give a few examples of the diagrams generated in our models.
If our aim is to apply the models to three-dimensional gravity, 
we should be able to assign three-dimensional volume to each diagram, 
and thus it is preferable that the diagrams can be regarded as collections only of tetrahedra. 
However, as we see in the examples below, 
there arise a lot of undesired diagrams. 
We will show in the next section 
that such undesired diagrams can be automatically excluded 
by taking specific associative algebras and modifying the form \eqref{C_def}, 
with an appropriate limit of parameters.

\subsubsection{Diagrams representing tetrahedral decompositions of manifolds} 
\label{ex_3dmfd}

\begin{figure}[htbp]
\centering 
\includegraphics[height = 4cm]{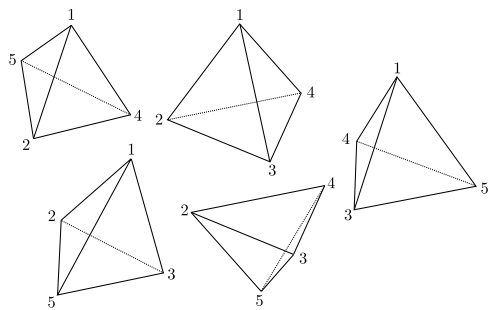}
\vspace{-5mm}
\begin{quote}
\caption{A decomposition of $S^3$ with five tetrahedra. 
The tetrahedra are glued together at their faces 
so that each of the points $1, \ldots, 5$ represents a single vertex. 
\label{5-cells}}
\end{quote}
\vspace{-10mm}
\end{figure}

\begin{figure}[htbp]
\begin{center}
\includegraphics[height = 4.0cm]{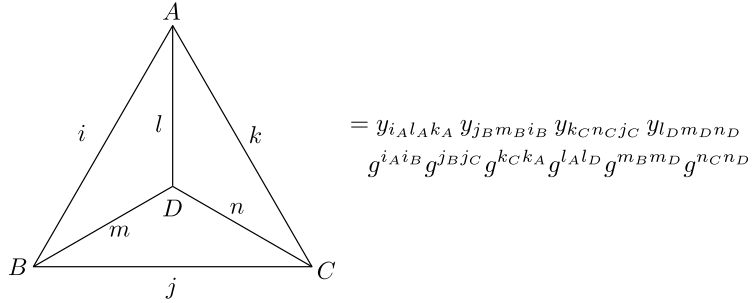}
\begin{quote}
\vspace{-5mm}
\caption{
A contraction of indices around a vertex. 
This represents a triangular decomposition of $S^2$ 
and gives the value $N$.  
}
\label{fig:s^2}
\end{quote}
\end{center}
\vspace{-6ex}
\end{figure}

First we consider a diagram which represents a tetrahedral decomposition 
of three-dimensional sphere $S^3$ 
(see Fig.~\ref{5-cells}). 
This is the boundary of the so-called 5-cell or a 4-simplex 
and can be constructed from five tetrahedra. 
Note that the diagram has ten triangles, ten 3-hinges and five vertices. 
All the index networks around vertices have the same topology 
and give triangular decompositions of $S^2$ as in Fig.~\ref{fig:s^2}. 
Thus, the neighborhood of each vertex is homeomorphic to $B^3$. 
Since every index network can be deformed to a single circle, 
the index function of each vertex takes the value $N$; 
$\zeta(v)=N=\mathcal{I}_{g(v)=0}$.  
Thus, the contribution from this diagram to the free energy is given by 
\begin{align}
 \frac{1}{S}\,\lambda^{10}\mu_3^{10} \, (\mathcal{I}_{g=0})^5
 =\frac{1}{S}\,\lambda^{10}\mu_3^{10} \, N^5, 
\end{align}
where $S$ is the symmetry factor of the diagram.

\begin{figure}[htbp]
\begin{center} 
\includegraphics[height = 4cm]{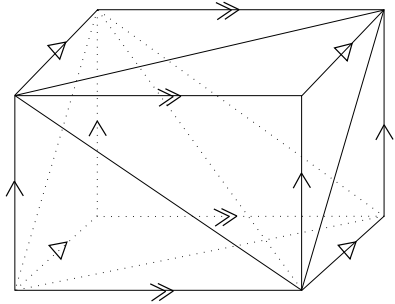}
\begin{quote}
\vspace{-3mm}
\caption{A tetrahedral decomposition of $T^3$.  
This is made by identifying the boundaries of a cuboid which consists of six tetrahedra.  
\label{T^3}}
\vspace{-5ex}
\end{quote}
\end{center}
\end{figure}

The next example is a diagram 
which represents a tetrahedral decomposition of three-dimensional torus $T^3$ 
(see Fig.~\ref{T^3}). 
The diagram has twelve triangles, four 4-hinges and three 6-hinges. 
It has only a single vertex due to the identification in the diagram. 
The index network around the vertex also represents $S^2$ 
as in the previous example for $S^3$.
Thus, the contribution from this diagram is given by
\begin{align}
 \frac{1}{S}\, \lambda^{12} \mu_4^4\, \mu_6^3 \, \mathcal{I}_{g=0}
 = \frac{1}{S}\, \lambda^{12} \mu_4^4\, \mu_6^3 \, N.
\end{align}

We can easily generalize the above results 
to such diagrams that represent tetrahedral decompositions of 
three-dimensional closed manifolds. 
Since the neighborhood of every vertex is homeomorphic to $B^3$, 
the contribution from such a diagram to the free energy is given by 
\begin{align}
 \frac{1}{S}\,\lambda^{s_2} \Bigl( \prod_{k \geq 2} \mu_k^{s_1^k} \Bigr) \, 
 (\mathcal{I}_{g=0})^{s_0} 
 =\frac{1}{S}\,\lambda^{s_2}
 \Bigl( \prod_{k \geq 2} \mu_k^{s_1^k} \Bigr) \, N^{s_0} ,
\label{wt_general}
\end{align}
where $s_2$, $s_1^k$ and $s_0$ represent, respectively, 
the number of triangles, $k$-hinges and vertices of the diagram. 
Note that since the topology of three-dimensional manifolds cannot be distinguished 
by $s_2$, $s_1^k$ and $s_0$ alone,%
\footnote{
We can read the number of tetrahedra, $s_3$, 
since the Euler characteristic of three-dimensional closed manifold is zero, 
$s_0-\sum_k s^k_1 +s_2 -s_3=0$.
} 
it can happen that topologically different manifolds 
give contributions of the same form. 
However, we in principle can distinguish the topology 
by carefully looking at the way of tetrahedral decompositions, 
although this is usually a tedious task.
Another way to examine the topology of diagrams 
is to evaluate a set of topological invariants of each diagram as in \cite{Chung:1993xr}.
This prescription will be further studied in our future paper \cite{fsu2}.

\subsubsection{Diagrams corresponding to pseudomanifolds} 
\label{ex_pmfd}

\begin{figure}[htbp]
\centering
\includegraphics[height = 4cm]{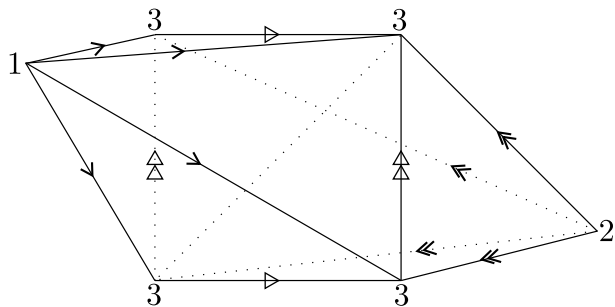}
\begin{quote}
\vspace{-3mm}
\caption{
A diagram which does not give a manifold. 
The neighborhood of of vertex 3 is $B^3$ 
but that of vertex 1 (and also 2) is a cone over $T^2$. }
\label{T^2cone}
\end{quote}
\vspace{-5mm}
\end{figure}

Our models also generate diagrams 
that have vertices whose neighborhoods are not three-dimensional ball $B^3$. 
One of such diagrams is depicted in  Fig.~\ref{T^2cone}, 
which consists of four tetrahedra, eight triangles, five edges and three vertices.  
The neighborhood of vertex 3 is homeomorphic to $B^3$, 
but that of vertex 1 (and also that of vertex 2) 
has the topology of cone over $T^2$. 
In fact, the index network around vertex 1 gives 
a polygonal decomposition of two-dimensional torus $T^2$.

One can check that the Euler characteristic of the diagram is not zero.
Thus, this diagram should not give a manifold 
(but still gives a pseudomanifold).  
The contribution from this diagram to the free energy can be evaluated to be  
\begin{align}
 \frac{1}{S}\, \lambda^8 \, \mu_4^3 \, \mu_6^2 \,
 \mathcal{I}_{g=0}\, (\mathcal{I}_{g=1})^2  ,
\end{align}
where $\mathcal{I}_{g=0}$ comes from the index network around vertex 3 
and equals $g_{ij}g^{ij}=N$. 
By contrast, two of $\mathcal{I}_{g=1}$ 
come from vertices 1 and 2, 
and have the value $p_{ij}g^{ij}=p^i_i$, 
which is the linear dimension of the center of  $\mathcal{A}$.

\subsubsection{Diagrams including singular cells} 
\label{ex_not_tetra}

\begin{figure}[htbp]
\centering \includegraphics[height = 3cm]{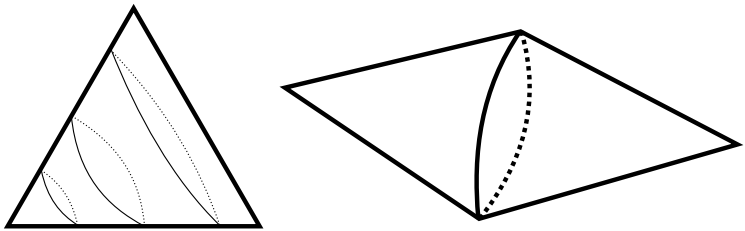}
\begin{quote}
\caption{Diagrams with singular cells. }
\label{3_triangles}
\end{quote}
\vspace{-5ex}
\end{figure}

There also arise diagrams 
which do not give tetrahedral decompositions. 
A few simple diagrams are depicted in Fig.~\ref{3_triangles} . 
Although they have the topology of $S^3$, 
it is not suitable to assign three-dimensional volume.    

\subsection{Strategy for the reduction to manifolds} 
\label{starategy}

We close this section 
by giving a strategy to choose the parameters in our models 
such that only tetrahedral decompositions of three-dimensional manifolds 
are generated as Feynman diagrams. 

As we will show in the proof of the theorem in subsection \ref{tetrahedron}, 
one can ensure a diagram to be a tetrahedral decomposition 
if the index network around every vertex is a {\em triangular} decomposition 
of two-dimensional surface. 
This condition can be realized 
by introducing a color structure to the models, 
as we will carry out in subsection \ref{intro_color}.

Furthermore,  
the manifoldness of the resulting diagrams can be ensured 
by appropriately choosing the defining associative algebra $\mathcal{A}$ 
such that the following two conditions are realized: 
(i) the number of vertices can be fixed 
by using free parameters in $\mathcal{A}$, 
and (ii)  $\mathcal{I}_{0}(v)\gg \mathcal{I}_{g}(v)$ for $g\geq 1$. 
In fact, due to the expression $\mathcal{F}(\gamma)=\prod_v \mathcal{I}_{g(v)}$ 
[see \eqref{index_fcn_prod_topological}],  
the dominant contributions come from 
the diagrams whose index networks all 
have the topology of two-sphere 
(and thus the neighborhood of every vertex has the topology of three-ball), 
namely, from the diagrams that represent (combinatorial) manifolds. 
If $\mathcal{A}$ does not have free parameters to fix the number of vertices, 
we extend $\mathcal{A}$ as needed. 
This extension will be carried out for matrix rings in subsection \ref{center}.

Note that our models also generate nonorientable diagrams. 
However, such diagrams always have an index network 
{\em not}\, homeomorphic to $S^2$ 
and thus are also decoupled in the program described in the previous paragraph.

\section{Matrix ring} 
\label{hm2mr}

In this section, we consider matrix rings as the defining associative algebras 
of the models. 
We show that such models can be constructed 
that generate only manifolds as Feynman diagrams, 
by introducing a color structure to the models 
and letting the associative algebras have centers 
whose dimensions play the role of free parameters 
(to count the number of vertices).

\subsection{The action and the Feynman rules for a matrix ring} 
\label{hm2mr_Feynman_rule}

Matrix ring $M_n(\mathbb{R})$ is the set of real-valued matrices 
of size $n$. 
This is an associative algebra with the same rules of 
addition, scalar product and multiplication as those of matrices. 
Note that $\mathcal{A}$ has the linear dimension $N=n^2$.    
Matrix ring is one of the simplest semisimple associative algebras 
because any semisimple associative algebra is isomorphic to 
a direct sum of matrix rings. 
In this section we analyze a model 
where $\mathcal{A}$ is set to be a matrix ring $M_n(\mathbb{R})$. 
We take its basis to be $\{e_{ab}\}$ ($a,b = 1, \dots , n$), 
where $e_{ab}$ is a matrix unit 
whose $(c,d)$ element is given by 
$(e_{ab})_{cd}=\delta_{ac} \,\delta_{bd}$. 
The structure constants can be read 
from the multiplication rule of matrices:
\begin{align}
 e_{ab} \times e_{cd} = \delta_{bc}\,e_{ad}
 = \delta^e_a \,\delta_{bc} \, \delta^f_d \, e_{ef}
 \equiv y_{abcd}^{\phantom{abcd}ef}e_{ef} .
\end{align}
We stress that the double index $(a,b)$ corresponds 
to the single index $i$ $(i=1,\ldots, N)$ 
in the previous section.%
\footnote{
The index $I$ in subsection \ref{subsec:general} thus becomes a quadruple index 
as $I=(i,j)=(a,b,c,d)$. 
} 
One can compute $y_{i_1 i_2\ldots i_k}=y_{a_1 b_1 a_2 b_2 \ldots a_k b_k}$ and $g^{ij}=g^{abcd}$ as
\begin{align}
y_{a_1 b_1 a_2 b_2 \ldots a_k b_k}
 = n\,\delta_{b_1 a_2}\delta_{b_2 a_3}\cdots \delta_{b_k a_1} ,\qquad
g^{abcd} = \frac{1}{n}\delta_{ad}\delta_{bc} .
\end{align}
By setting the tensor $C^{ijklmn}$ as in \eqref{C_def}:
\begin{align}
 C^{a_1 b_1 c_1 d_1 a_2 b_2 c_2 d_2 a_3 b_3 c_3 d_3} 
 = \frac{1}{n^3}\, \delta^{d_1 a_2} \delta^{c_1 b_2} \delta^{d_2 a_3}
 \delta^{c_2 b_3} \delta^{d_3 a_1} \delta^{c_3 b_1},
\label{C_matrix}
\end{align}
the action \eqref{hm2_action} has the form 
\begin{align}
 S = \frac{1}{2} A_{abcd} B^{abcd} - \frac{\lambda}{6n^3} A_{bacd} A_{dcef} A_{feab} 
 - \sum_{k \geq 2} \frac{n^2 \mu_k}{2k} B^{a_1 a_2 b_2 b_1} B^{a_2 a_3 b_3 b_2}
 \cdots B^{a_k a_1 b_1 b_k} ,
\label{hm2mr_action}
\end{align}
where $A$ and $B$ satisfy the following relations 
because of the symmetry property \eqref{sym_AandB}:
\begin{align}
 A_{abcd}=A_{cdab}, \quad B^{abcd}=B^{cdab}.
\label{matrix_sym}
\end{align}

\begin{figure}[htbp]
\begin{center}
\includegraphics[height = 5cm]{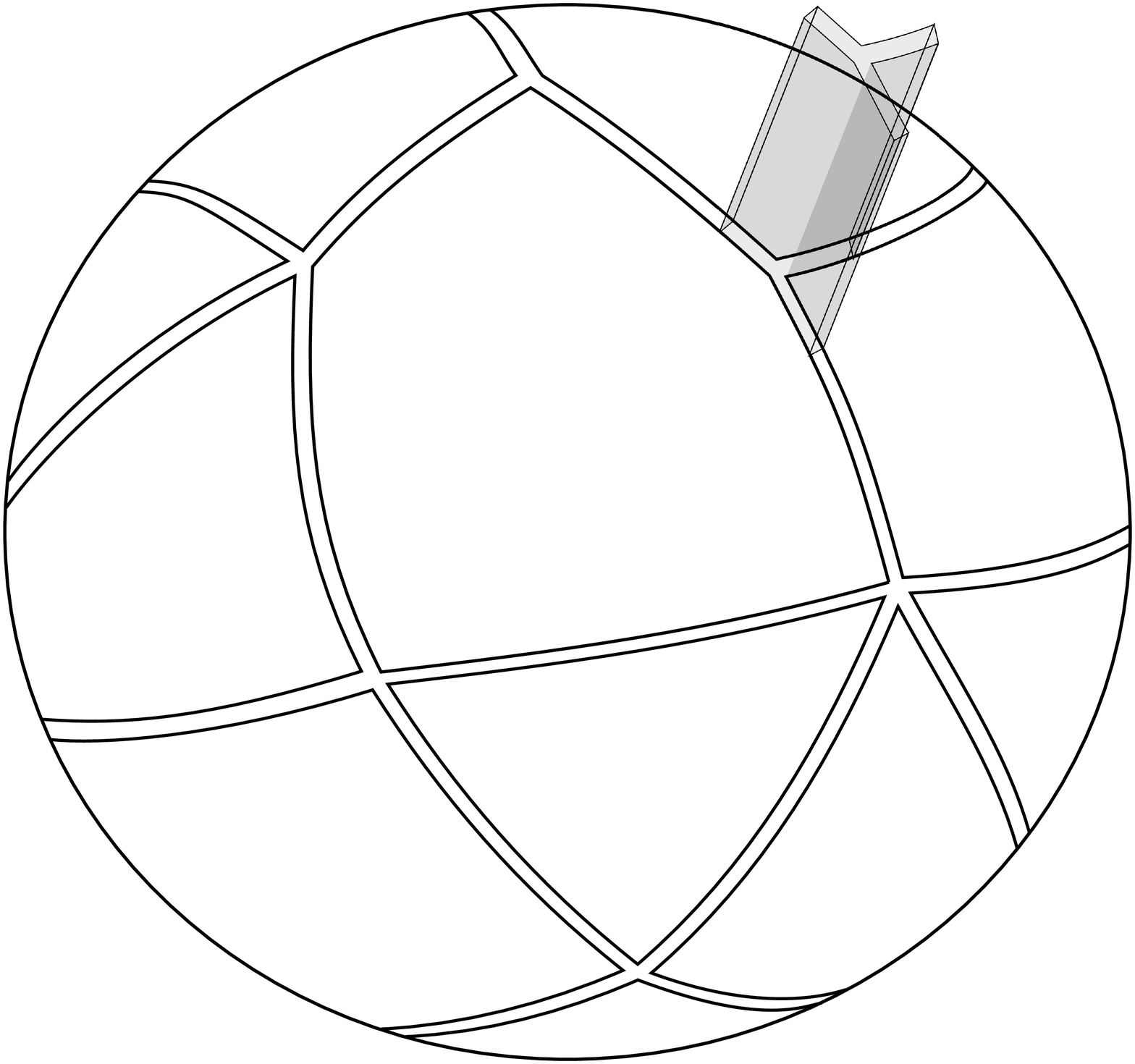}
\begin{quote}
\vspace{-5mm}
\caption{
Index network around a vertex $v$ 
when $\mathcal{A}$ is a matrix ring $M_n(\mathbb{R})$. 
Everything is the same as Fig.~\ref{fig:index_network} 
except that the index lines are now double lines. 
The index network represents a closed oriented surface 
(not necessarily a sphere).
}
\label{fig:index_network_matrix}
\end{quote}
\end{center}
\vspace{-6ex}
\end{figure}

The Feynman rules for the action \eqref{hm2mr_action} 
can be expressed with quadruple lines as follows:
\begin{align}
 {\rm propagator}&:\ \begin{array}{l} \includegraphics[height = 1.5cm]{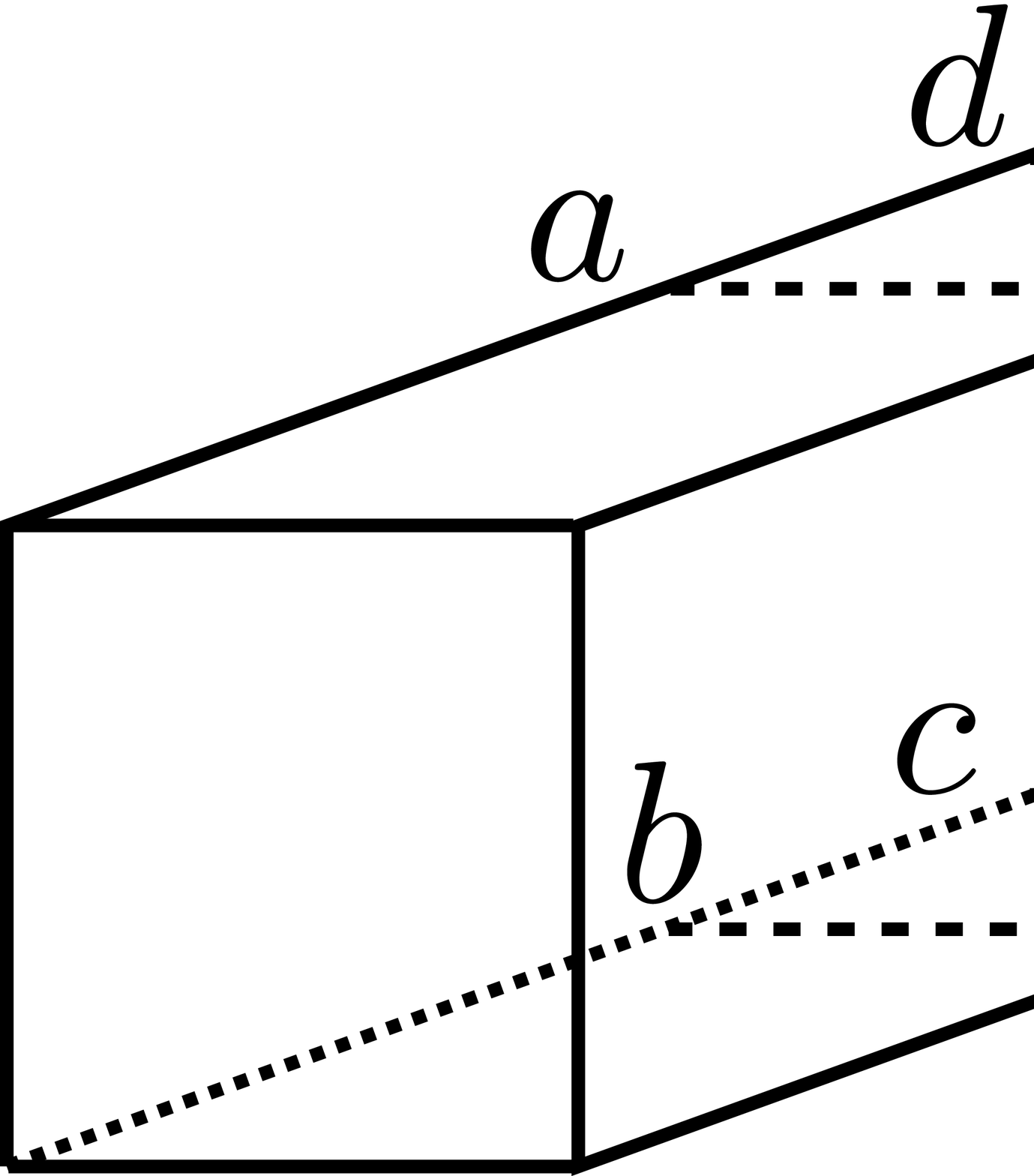} \end{array} 
 + \begin{array}{l} \includegraphics[height = 1.5cm]{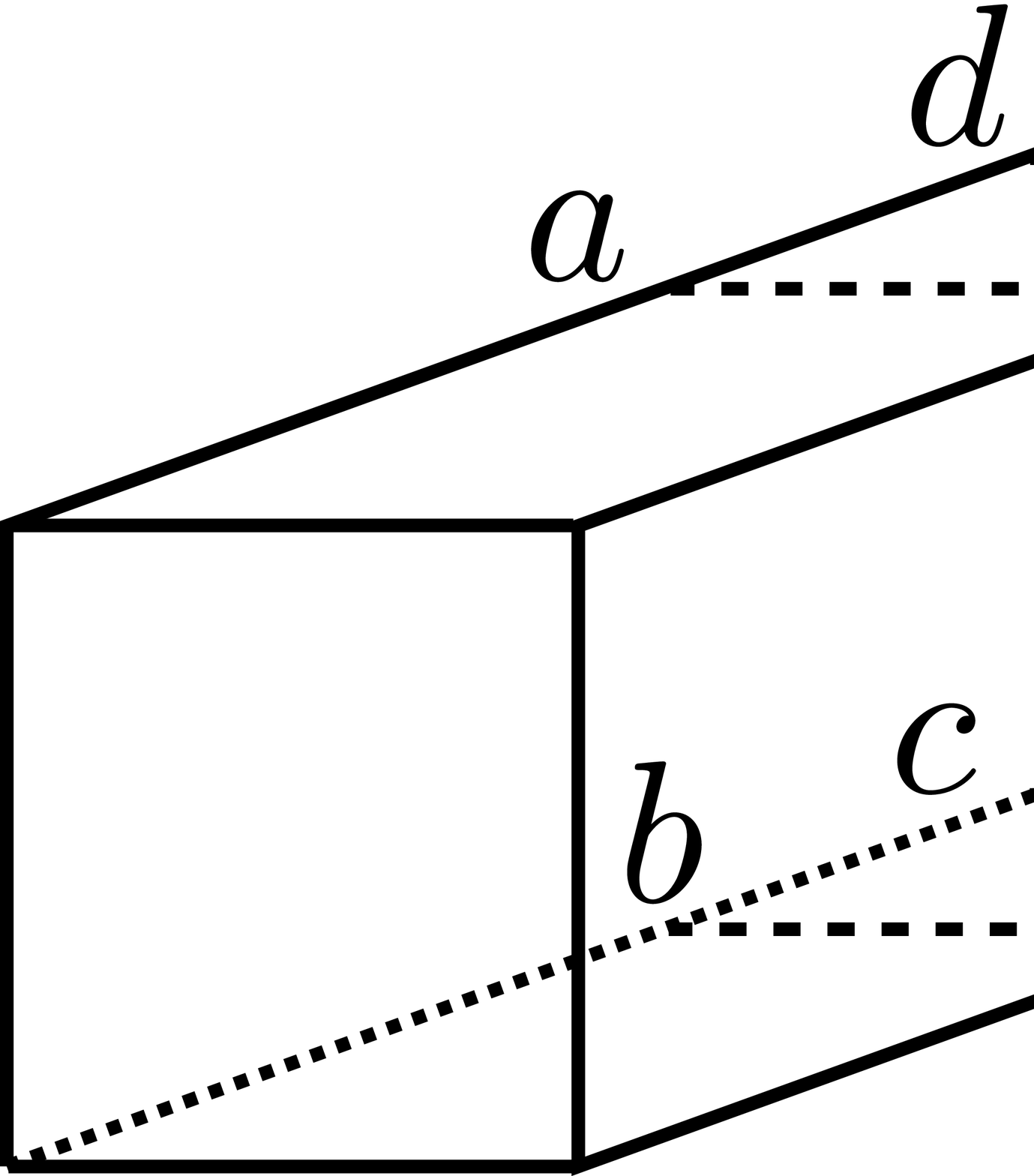} \end{array} \nonumber \\ 
 & \qquad\qquad \sim \langle A_{abcd}B^{efgh} \rangle 
 = \delta_a^{\phantom{a}e} \delta_{b}^{\phantom{b}f}
 \delta_c^{\phantom{c}g} \delta_d^{\phantom{d}h} 
 + \delta_a^{\phantom{a}g} \delta_{b}^{\phantom{b}h}
 \delta_c^{\phantom{c}e} \delta_d^{\phantom{d}f} ,
\label{hm2mr_propagator} 
\\
 {\rm triangle}&:\ \begin{array}{l} \includegraphics[width = 3.3cm]{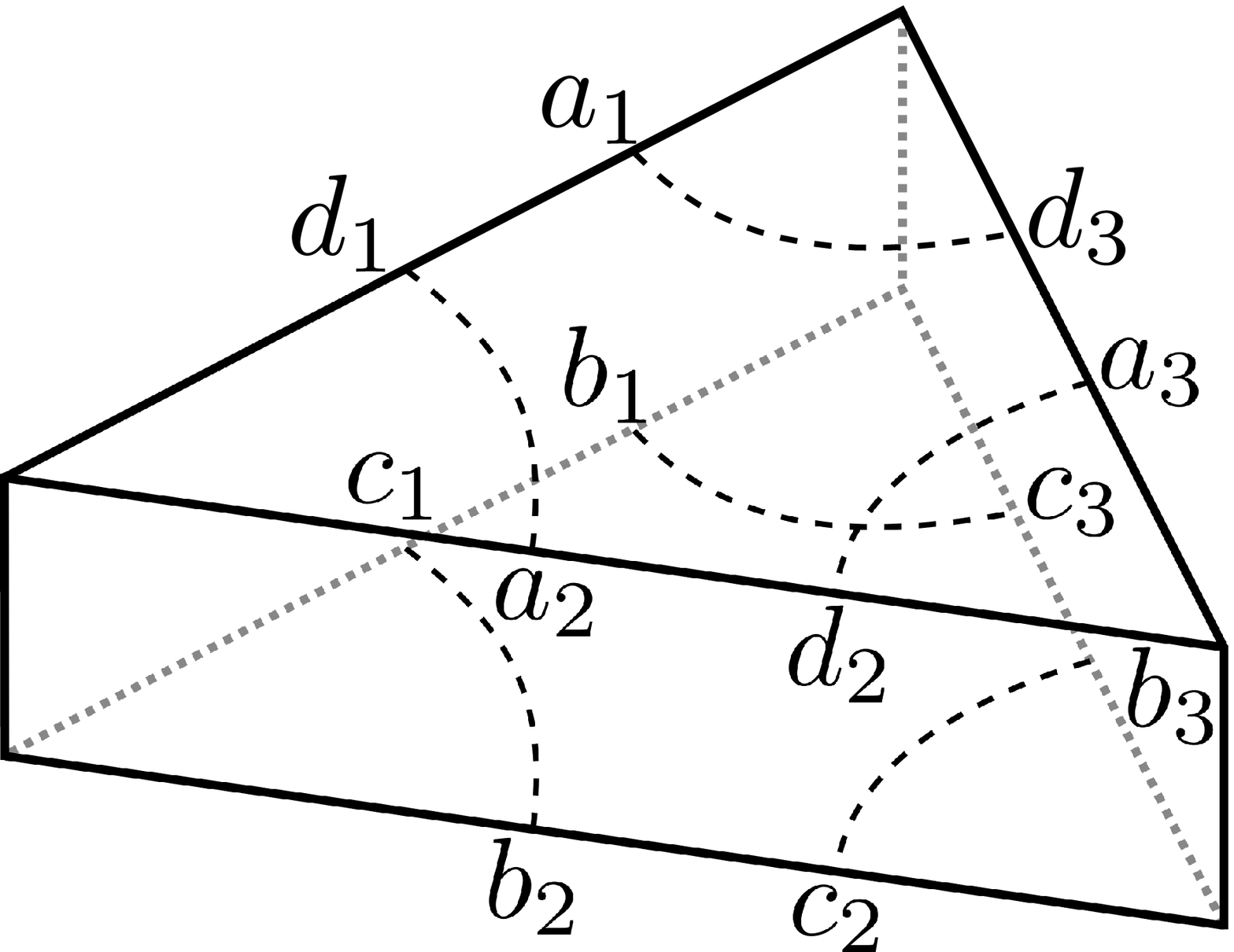} \end{array} 
 \sim \frac{\lambda}{n^3} \,\delta^{d_1a_2} \delta^{c_1b_2} \ldots
 \delta^{d_3a_1} \delta^{c_3b_1} ,
\label{hm2mr_triangle} 
\\
 \mbox{$k$-hinge}&:\ \begin{array}{l} \includegraphics[width = 2.6cm]{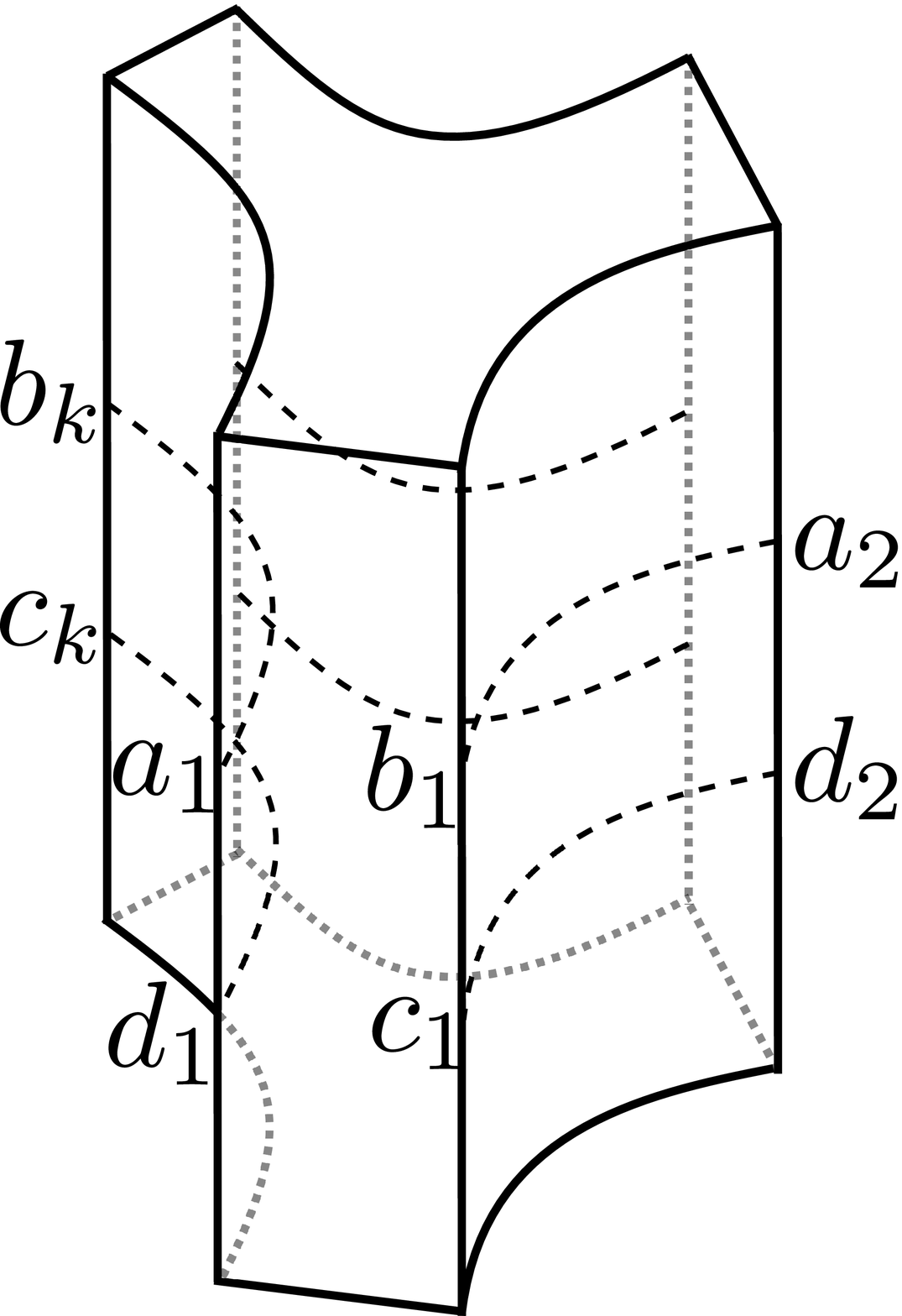} \end{array} 
 \sim n^2 \mu_k \, \delta_{b_1 a_2} \delta_{c_1d_2} \ldots \delta_{b_ka_1} \delta_{c_kd_1} .
\label{hm2mr_hinge}
\end{align}
Note that each of the index lines in \eqref{hm2_propagator}--\eqref{hm2_hinge} 
becomes a double line. 
Moreover, the index line dose not have branch points in this case
due to the index structure of hinges 
[see \eqref{hm2mr_hinge}]. 
Thus, as depicted in Fig.~\ref{fig:index_network_matrix},  
the identification of the index network  
with a polygonal decomposition of two-dimensional surface  
can be done automatically (and uniquely)%
\footnote{
Each polygonal face is specified as the region 
bounded by a closed loop for index $a$.
} 
[although this identification can also be carried out uniquely 
even when $\mathcal{A}$ is not a matrix ring, 
as argued in the first comment 
following \eqref{index_fcn_prod}]. 

The contribution from each index network to the free energy can be calculated 
just as in the standard matrix model. 
To see this, we first note that each polygon gives a factor of $n$ 
because each index loop 
(i.e.\ the index contraction with respect to one of the double index) 
gives 
$\delta^a_a=n$. 
We also see from the coefficients in \eqref{hm2mr_triangle} and \eqref{hm2mr_hinge} 
that each segment in the polygonal decomposition 
gives $n^{-1}$ (one-third contribution from a triangle) 
and each junction gives $n$ (one-half contribution from a hinge). 
In total, the contribution from the index network around vertex $v$ in the original diagram 
is given by 
$n^{\#({\rm polygon})-\#({\rm segment})+\#({\rm junction})}
=n^{2-2 g(v)}$, 
where $g(v)$ is the genus of the index network around $v$. 
One can easily see that an insertion of the projector $p_{ij}$, \eqref{proj_p}, into the diagram 
corresponds to attaching a handle to the index network (as in \cite{Fukuma:1993hy})
and decreases the power of $n$ by two.

\subsection{Color structure} 
\label{intro_color}

In subsection \ref{ex_not_tetra} 
we argued that undesired diagrams appear in our models. 
In this subsection 
we show that they can be excluded 
by introducing a ``color structure'' to our models.

\begin{figure}[htbp]
\centering \includegraphics[height = 3.5cm]{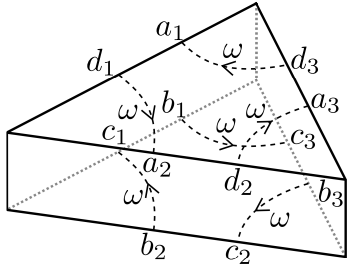}
\vspace{-3mm}
\begin{quote}\caption{
Triangles with the color structure. 
$\omega$ has the value $\omega^{da}$ 
when it is inserted into the index line from $d$ to $a$. 
Note that $\omega^{bc}=(\omega^{-1})^{cb}$. 
 }
\label{fig:triangle_color}
\end{quote}
\vspace{-5ex}
\end{figure}

Let the size $n$ of matrices be a multiple of three, $n=3m$. 
We then modify the tensor \eqref{C_matrix}
to 
\begin{align}
 C^{a_1 b_1 c_1 d_1 a_2 b_2 c_2 d_2 a_3 b_3 c_3 d_3} 
 = \frac{1}{n^3}\, \omega^{d_1 a_2} \omega^{b_2 c_1}
 \omega^{d_2 a_3} \omega^{b_3 c_2} \omega^{d_3 a_1} \omega^{b_1 c_3},  
\label{C_color}
\end{align}
where $\omega$ is a permutation matrix of the form
\begin{align}
\omega \equiv 
\begin{pmatrix} 0&1_m&0 \\ 0&0&1_m \\ 1_m&0&0 \end{pmatrix},
& \qquad 1_m : m\times m \, \text{unit matrix}.
\end{align}
This modification%
\footnote{
Although we only discuss the case $\mathcal{A}=M_{3m}(\mathbb{R})$, 
we can also introduce the color structure to other algebras 
by taking the tensor product of the form 
$
 \mathcal{R} = (\mathcal{A} \otimes M_{3}(\mathbb{R}))
 \otimes (\mathcal{\bar{A}} \otimes \overline{M_{3}(\mathbb{R})}). 
$
Note that $M_{m}(\mathbb{R})\otimes M_3(\mathbb{R})=M_{3m}(\mathbb{R})$. 
Then, the variables $A$ and $B$ are expressed as 
$A_{ij(abcd)}=A_{ji(cd ab)}$ and $B^{ij(abcd)}=B^{ji(cdab)}$, 
and the action has the form
\begin{align}
 S&=\frac{1}{2}\,A_{ij(abcd)}\,B^{ij(abcd)}
\nn\\
 &~~~~-\frac{\lambda}{6\cdot 3^3}\,A_{ij(a_1 b_1 c_1 d_1)}\,g^{jk}
 A_{kl(a_2 b_2 c_2 d_2)}\,g^{lm}
 A_{mn(a_3 b_3 c_3 d_3)}\,g^{ni}\,
 \omega^{d_1 a_2} \omega^{b_2 c_1}
 \omega^{d_2 a_3} \omega^{b_3 c_2} \omega^{d_3 a_1} \omega^{b_1 c_3}\nn\\
 &~~~~-\sum_{k\geq 2} \frac{3^2 \mu_k}{2k}\,B^{i_1j_1(a_1 a_2 b_2 b_1)}\cdots
 B^{i_k j_k(a_k a_1 b_1 b_k)}\,
 y_{i_1\ldots i_k}\,y_{j_k\ldots j_1}.
\nn 
\end{align}
} 
corresponds to inserting $\omega$ and $\omega^{-1}=\omega^T$ in a pair 
into two index lines on every segment in each index network 
(see Fig.~\ref{fig:triangle_color}).
Note that 
only $\omega$ (not $\omega^{-1}$) are accumulated 
when following the arrows in each index line. 
Thus, the value of a closed index loop forming $\ell$-gon 
changes from $\mathrm{tr} \,1_{n}=n=3 m$ to
\begin{align}
\mathrm{tr}(\omega^\ell) = \left\{ \begin{array}{l}
n \phantom{0}\hspace{1em} (\ell = 0 \mod 3) \\
0 \phantom{n}\hspace{1em} (\ell \neq 0 \mod 3) .
\end{array}\right.
\end{align}
We thus see that 
the index function of a diagram gives a nonvanishing value 
only when the index network around every vertex has a polygonal decomposition 
where the number of segments of each polygon 
is a multiple of three.

Note that such polygonal decompositions with nonvanishing index functions 
have the following dependence on the coupling constants. 
Suppose that the index network around vertex $v$ has 
$t_2(v)$ polygons, $t_1(v)$ segments 
and $t_0(v)$ junctions.  
Here, $t_2(v) =\sum_\ell t^\ell_2(v)$  
with $t^\ell_2(v)$ the number of $\ell$-gons, 
and $t_0(v) =\sum_k t^k_0(v)$  
with $t_0^{k}(v)$ the number of $k$-junctions.  
It is easy to see that 
the function $d(v) \equiv 2t_1(v) - 3t_2(v)$ 
can be expressed as $d(v) = \sum_\ell (\ell-3)\,t^\ell_2(v)$. 
Thus $d(v)$ is nonnegative for the $C$'s in \eqref{C_color} 
because monogons and digons 
are excluded due to the color structure 
[i.e., $t^{\ell =1}_2(v)=t^{\ell=2}_2(v)=0$]. 
Recalling that the contribution from each diagram is given by 
\begin{align}
 \frac{1}{S} \,\lambda^{s_2} \,
 \Bigl(\, \prod_{k \geq 2} \mu_k^{s_1^k} \Bigr)
 \prod_{v:\,\text{vertex}}  n^{2-2g(v)}, 
\label{wt_color}
\end{align}
and noting that the identification rule of the polygonal decompositions 
gives the relations
\begin{align}
 s_2= \frac13 \sum_v t_1(v), \quad 
 s_1^k =  \frac12 \sum_v t_0^{k}(v), 
\end{align}
we find another expression of  \eqref{wt_color}:
\begin{align}
 \frac{1}{S}\,\lambda^{s_2} \,
 \Bigl(\, \prod_{k \geq 2} \mu_k^{s_1^k} \Bigr)
 \prod_{v:\,\text{vertex}} n^{2-2g(v)}  
 = \frac{1}{S}\,\prod_{v:\,\text{vertex}}
 \biggl[
 \Bigl[\, \prod_{k \geq 2} (\lambda^2 \mu_k)^{\frac{1}{2} t_0^{k}(v)} \Bigr]
 \Bigl( \frac{n}{\lambda} \Bigr)^{2-2g(v)} 
 \Bigl( \frac{1}{\lambda} \Bigr)^{\frac{1}{3}d(v)} 
 \biggr].
\label{hm2mr_colored_diag_value} 
\end{align}
Therefore, 
if we expand  the free energy around $\lambda=\infty$ 
with $\lambda^2\, \mu_k$ and $n/\lambda$ being fixed, 
the leading contribution comes from 
such diagrams that satisfy $d(v)=0$ 
for every vertex $v$, 
namely, from the diagrams 
where every index network forms a {\em triangular} decomposition.

\subsection{Counting the number of vertices} 
\label{center}

One may think from \eqref{hm2mr_colored_diag_value} 
that it would be possible by taking a limit $n/\lambda \to \infty$ 
to single out the diagrams 
where the index networks are all homeomorphic to two-sphere $S^2$.
However, this is not the case. 
For example, 
let us consider a diagram which includes an index network forming a two-torus $T^2$. 
Since the index network gives the contribution of $(n/\lambda)^0 = 1$,
we cannot distinguish a diagram 
whose vertices all give index networks homeomorphic to $S^2$ 
from a diagram which has the same number of such vertices 
whose index networks are homeomorphic to $S^2$ 
but also has extra vertices whose index networks are homeomorphic to $T^2$, 
because the contributions from the two diagrams to the free energy 
have the same form.

This problem comes from the fact that we cannot control the number of vertices 
only with the coupling constants 
existing in the model with $\mathcal{A}=M_n(\mathbb{R})$. 
However, this can be remedied by setting  the algebra $\mathcal{A}$ 
to be the direct sum of $K$ copies of 
matrix ring $\mathcal{A}_0=M_n(\mathbb{R})$,%
\footnote{
The following prescription to count the number of vertices 
can be directly applied to any associative algebras $\mathcal{A}_0$. 
} 
\begin{align}
 \mathcal{A} =
 \underbrace{\mathcal{A}_0 \oplus \cdots \oplus \mathcal{A}_0}_{K~{\rm copies}}
 = K \mathcal{A}_0. 
\end{align}
In fact, 
the index function of a diagram with $s_0$ vertices 
becomes proportional  to $K^{s_0}$ 
since the index network around each vertex 
gives a factor of $K$ independently, 
and thus \eqref{hm2mr_colored_diag_value}  changes to%
\footnote{
Note that $K$ equals the linear dimension of $Z(\mathcal{A})$. 
} 
\begin{align}
 \frac{1}{S}\,\prod_{v: \,\text{vertex}} \biggl[
 K \Bigl[\prod_{k \geq 2} (\lambda^2 \mu_k)^{\frac{1}{2} t_0^{k}(v)}\Bigr] 
 \Bigl( \frac{n}{\lambda} \Bigr)^{2-2g(v)} 
 \Bigl( \frac{1}{\lambda} \Bigr)^{\frac{1}{3}d(v)}  \biggr].
\end{align}
Therefore, we can single out the diagrams 
where every index network is homeomorphic to  $S^2$, 
by picking out only the diagrams whose index function 
gives the values with the same power of $K$ as that of $n^2$.

\subsection{Reduction to manifolds} 
\label{tetrahedron}

Combining the results in subsections \ref{intro_color} and \ref{center},  
we can reduce the set of possible diagrams to those 
whose index networks all give triangular decompositions of two-sphere $S^2$. 
We then can apply the following theorem 
to conclude that these diagrams represent tetrahedral decompositions 
of three-dimensional manifolds:

\noindent
{\bf Theorem.}
{\it 
Assume that the index network around every vertex in diagram $\gamma$ 
gives a triangular decomposition of two-sphere. 
Then, $\gamma$ represents a tetrahedral decomposition 
of a three-dimensional manifold. 
}

\begin{proof}
We label the vertices, triangles and hinges of diagram $\gamma$ 
as $v$ $(=1,2,3,\ldots)$, $f$ $(=i,j,k,\ldots)$ 
and $h$ $(=A,B,C,\ldots)$, respectively. 
Let $\mathcal{T}_v$ denote the index network around vertex $v$, 
which we assume to have a form of triangular decomposition of two-sphere.  
Note that every corner of a triangle in $\gamma$ corresponds 
to a segment of the index network around some vertex 
(see Fig.~\ref{proof_fig}). 
\begin{figure}[htbp]
\centering \includegraphics[height = 5cm]{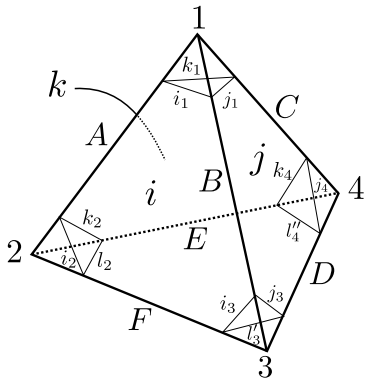}
\vspace{-3mm}
\begin{quote}\caption{
A part of diagram $\gamma$. 
The triangle $(i_1,j_1,k_1)$ is a part of the index network around vertex 1, 
which has a form of triangular decomposition. }
\label{proof_fig}
\end{quote}
\vspace{-5ex}
\end{figure}
We denote by $f_v$ the segment 
which is lying on triangle $f$ and is placed in the corner at vertex $v$. 

We choose a vertex (say $v=1$) 
and focus on an ``index triangle'' formed by three segments $i_1, j_1, k_1$ 
in $\mathcal{T}_1$. 
Here, $i,j,k$ are the triangles 
on which the three segments live. 
Since all the edges of each triangle are attached to hinges, 
there are hinges $A=(12)$, $B=(13)$, $C=(14)$, $D=(34)$, $E= (42)$, $F=(23)$ 
as in Fig.~\ref{proof_fig}.%
\footnote{
Note that some of vertices $1,2,3,4$ may represent the same vertex 
because the index triangles around them 
may belong to the same connected component 
of an index network. 
} 
As is depicted there, 
the three index triangles $(i_2,k_2,l_2)$, $(i_3,j_3,l'_3)$ and $(j_4,k_4,l''_4)$ 
ensure the existence of the corresponding triangles 
$l$, $l'$ and $l''$, respectively. 
We are now going to give a detailed description of these triangles 
and show that they all coincide, $l=l'=l''$.

We first take a look at hinge $A=(12)$.  
We assume that $2\to 1$ is the positive direction of hinge $A$ 
and label the triangles 
such that triangle $k$ is to the immediate left of $i$ 
when seen from vertex 1 
(see Fig.~\ref{proof_fig}). 
This means that triangle $k$ is to the immediate right of $i$ 
when seen from vertex 2, 
so that $i_2$ and $k_2$ are two segments of an index triangle around vertex 2, 
which will be complemented by the third segment $l_2$ as in Fig.~\ref{proof_fig}.
The triangle $l$ on which the segment $l_2$ lives 
is glued to triangle $i$ along hinge $F=(23)$, 
and must be to the immediate left of $i$ 
when seen from vertex 2 in the direction of $F$.

We repeat the same argument for hinge $B=(13)$. 
There, triangle $j$ is to the immediate right of $i$ when seen from vertex 1.  
This means that triangle $j$ is to the immediate left of $i$ when seen from vertex 3, 
so that $i_3$ and $j_3$ are two segments of an index triangle around vertex 3, 
which will be complemented by the third segment $l'_3$ as in Fig.~\ref{proof_fig}.
The triangle $l'$ on which the segment $l'_3$ lives 
is glued to triangle $i$ along hinge $F=(23)$, 
and must be to the immediate right of $i$ 
when seen from vertex 3 in the direction of $F$. 
However, this means that $l'$ is to the immediate left of $i$ 
when seen from vertex 2, 
and thus two triangles $l$ and $l'$ must be the same.

The same argument can also be made for hinge $C=(14)$, 
and we obtain $l=l'=l''$, 
from which we see that there exists a tetrahedron 
surrounded by four triangles $i$, $j$, $k$, $l$.
By repeating the same arguments for all the index triangles around every vertex, 
we conclude that diagram $\gamma$ gives a tetrahedral decomposition. 
Furthermore, since the index network around every vertex 
represents a triangular decomposition of $S^2$, 
the neighborhood of every vertex is homeomorphic to $B^3$. 
Therefore, the diagram $\gamma$ gives a tetrahedral decomposition 
of a three-dimensional manifold. 
\end{proof}

\subsection{Three-dimensional gravity}
\label{3d_gr}

We have shown that a class of our models allow us to single out the diagrams 
which represent tetrahedral decompositions of three-dimensional manifolds. 
Such models can be used to define discretized three-dimensional Euclidean gravity. 
In fact, we only need to  follow the arguments 
given in \cite{Weingarten:1982mg, Ambjorn:1990ge, Sasakura:1990fs}.

The action of three-dimensional Euclidean gravity is given by
\begin{align}
  S_0=-\kappa_0 \int\! d^3 x \sqrt{g}\, R + \Lambda_0 \int\! d^3 x \sqrt{g}, 
\end{align}
where $\kappa_0$ corresponds to the bare gravitational coupling 
and $\Lambda_0$ to the bare cosmological constant.  
This can be discretized by using regular tetrahedra with fixed spacing $a$ as 
\begin{align}
 S_{\mathrm{EH}}= -4\pi\kappa_0 a \,s_0\,
+ \Bigl[\frac{\sqrt{2}a^3}{12} \Lambda_0
 - 4\pi \kappa_0 a \Bigl(1-\frac{3\theta}{\pi}\Bigr) \Bigr]s_3.
\end{align}
Here, $s_0$ and $s_3$ denote the number of vertices and tetrahedra, respectively, 
and $\theta\equiv\arccos(1/3)$ is the angle 
between two neighboring triangles in a regular tetrahedron. 
The free energy of this action is then given by
\begin{align}
\log Z_{\mathrm{EH}}
 &= \sum_{\mathrm{config.}} \frac{1}{S}\, e^{-S_{\mathrm{EH}}} \nn\\
 &=\sum_{\mathrm{config.}} \frac{1}{S} \, (e^{4\pi\kappa_0 a})^{s_0}\, 
 \bigl(e^{-\frac{\sqrt{2}a^3}{12} \Lambda_0
 + 4\pi \kappa_0 a \bigl(1-\frac{3\theta}{\pi}\bigr) }\bigr)^{s_3}, 
\label{free_energy_EH}
\end{align}
where $S$ is the symmetry factor.

In our models, on the other hand, 
each diagram representing a tetrahedral decomposition 
contributes to the free energy as 
\begin{align}
 \frac{1}{S}\,\lambda^{s_2} \mu^{s_1} N^{s_0},  
\end{align}
Here we have set $\mu_k\equiv \mu$ $(\forall k\geq 2)$. 
Since the relations $s_2=2 s_3$ and $s_1=s_0 + s_3$ hold  
for tetrahedral decompositions of a three-dimensional manifold, 
the contribution takes the form
\begin{align}
 \frac{1}{S}\, (\mu N)^{s_0} (\lambda^2\mu)^{s_3} .  
\label{free_energy_FSU}
\end{align}
Comparing \eqref{free_energy_EH} and \eqref{free_energy_FSU}, 
we obtain the relations between the coupling constants 
of the two models, 
\begin{align}
 \mu N= e^{4\pi\kappa_0 a}, \qquad
 \lambda^2\mu = e^{-\frac{\sqrt{2}a^3}{12} \Lambda_0 
 + 4\pi \kappa_0 a \bigl(1-\frac{3\theta}{\pi}\bigr) } .
\end{align}

\subsection{Duality} 
\label{mr_duality}

We conclude this section 
by commenting that there exists a novel strong-weak duality 
which interchanges the roles of triangles and hinges 
when $\mathcal{A}$ is a matrix ring. 
We expect this duality to play an important role 
when we further study the analytic properties of the models 
in the future.  

We first recall that one has two choices 
when introducing a structure of associative algebra 
to the tensor product of linear spaces, 
$\mathcal{R}=\mathcal{A}\otimes\bar{\mathcal{A}}$ 
[see \eqref{endomorphisms}]. 
The first is the algebra structure as the tensor product 
of two associative algebras $\mathcal{A}$ and $\bar{\mathcal{A}}$. 
This is the structure we have used exclusively so far,  
and gives the multiplication \eqref{Yyy} (denoted by $\times$), 
which can also be written as 
\begin{align}
 (B_1\times B_2)^{ij} \equiv (B_1\times B_2)^i_{\phantom{i}k}\, \sigma^{kj}
 =B_1^{kl} B_2^{mn}  y_{km}^{\phantom{km} i} y_{nl}^{\phantom{nl} j} 
 \quad \text{for} ~~ B_1 , B_2 \in  \mathcal{R} .
\end{align}
The second is the algebra structure as the set of endomorphisms of $\mathcal{A}$\,; 
$\mathcal{R}={\rm End}\,\mathcal{A}$. 
The multiplication is defined as the composition of two linear operators 
acting on $\mathcal{A}$ 
and will be denoted by dot ``$\,\cdot\,$'': 
\begin{align}
 B_1 \cdot B_2 = (B_1 \cdot B_2)^i_{\phantom{i}j} e_i \otimes \bar{e}^j  
 \equiv(B_1)^i_{\phantom{i}k} (B_2)^k_{\phantom{k}j} e_i \otimes \bar{e}^j 
 \quad \text{for} ~~ B_1 , B_2 \in  \mathcal{R} , 
\label{R_cdot}
\end{align}
which can also be written as
\begin{align}
 (B_1\cdot B_2)^{ij} \equiv (B_1\cdot B_2)^i_{\phantom{i}k}\,\sigma^{kj}
 = B_1^{ik} \sigma_{kl} B_2^{lj} .
\end{align}
We will show that there is a duality between the two algebra structures 
when $\mathcal{A}$ is a matrix ring.

We first set $\sigma_{ij}=g_{ij}$. 
This is possible because $\sigma$ can be chosen in an arbitrary way 
[see a comment following \eqref{hm2_action}]. 
Then, when $\mathcal{A}=M_{n}(\mathbb{R})$, 
the multiplications are represented as 
\begin{align}
(B_1 \times B_2)^{abcd} &= B_1^{aefd}B_2^{ebcf} ,
\label{prod_cross}
\\
(B_1 \cdot B_2)^{abcd} &=B_1^{abef} g_{efgh} B_2^{ghcd} = n\, B_1^{abef}B_2^{fecd}.
\label{prod_dot}
\end{align}
We now introduce the dual variables $\tilde{B}$ to $B$ as 
\begin{align}
 \tilde{B}^{abcd} \equiv B^{bcda}, 
\end{align}
which satisfy the symmetry property $\tilde{B}^{abcd} = \tilde{B}^{cdab}$ 
due to \eqref{matrix_sym}. 
Then one can easily show from \eqref{prod_cross} and \eqref{prod_dot} 
that the two multiplications are interchanged for the dual variables:
\begin{align}
 (B_1 \times B_2)^{abcd} = \frac{1}{n} (\tilde{B}_2 \cdot \tilde{B}_1)^{bcda}, \qquad
 (B_1 \cdot B_2)^{abcd} = n (\tilde{B}_1 \times \tilde{B}_2)^{bcda}.
\end{align}
We further introduce the variables $\tilde{A}$ dual to $A$ as
\begin{align}
 \tilde{A}_{abcd} \equiv A_{bcda}\,\bigl(=\tilde{A}_{cdab}\bigr). 
\end{align}
Then the action \eqref{hm2mr_action} can be rewritten in terms of 
the dual variables $\tilde{A}$ and $\tilde{B}$ to the form
\begin{align}
 S = \frac{1}{2} \tilde A_{abcd} \tilde B^{abcd}
 - \frac{\lambda}{6n^3} \tilde A_{abcd} \tilde A_{befc} \tilde A_{eadf} 
 - \sum_{k \geq 2} \frac{n^2 \mu_k}{2k} \tilde B^{a_1 b_1 b_2 a_2}
 \tilde B^{a_2 b_2 b_3 a_3} \ldots \tilde{B}^{a_k b_k b_1 a_1} .
\label{action_dual}
\end{align}
Note that the way to contract the indices of $\tilde{A}$ (or $\tilde{B}$) 
in the dual action \eqref{action_dual} 
is the same as that of $B$ (or $A$) in the original action \eqref{hm2mr_action}.  
This means that a triangle for the original variables, \eqref{hm2mr_triangle}, 
now plays the role of a 3-hinge for the dual variables, 
and a $k$-hinge for the original variables, \eqref{hm2mr_hinge},  
plays the role of a $k$-gon for the dual variables. 
We thus find that the action \eqref{action_dual} for the dual variables 
generates the dual diagrams to the original ones, 
consisting of $3$-hinges (dual to original triangles) 
and polygons (dual to original hinges).%
\footnote{
The duality between the two actions will become more symmetric 
if one allows $k$-gons to appear in the original action for all $k \geq 2$ . 
} 
Note that the large $N$ limit in \eqref{hm2mr_colored_diag_value} 
($n\to\infty$ with $\lambda^2 \mu_k$ and $n/\lambda$ being fixed) 
gives $\lambda\to \infty$ and $\mu_k=\mu\to 0$. 
Since $\lambda$ and $\mu$ are interchanged in the duality transformation, 
one sees that this duality is actually a strong-weak duality.

\section{Group ring} 
\label{hm2gr}

In this section, 
we investigate the models 
where $\mathcal{A}$ is set to be a group ring $\mathbb{R}[G]$, 
and demonstrate how the models depend on details of the group structure of $G$. 
We assume that $G$ is a finite group with order $|G|$ 
in order to avoid introducing regularizations, 
although most of the relations below can be applied 
to continuous compact groups.

\subsection{Action for a group ring}
\label{group_ring_model}

Group ring $\mathbb{R}[G]$ is an associative algebra 
linearly spanned by the elements of $G$, 
$\mathbb{R}[G]=\bigoplus_{x\in G}\mathbb{R} e_x$, 
with multiplication rule determined by that of group $G$, 
\begin{align}
e_x \times e_y = e_{xy} .
\end{align}
The structure constants $y_{x,y}^{\phantom{x,y}z}$ are then given by 
\begin{align}
y_{x,y}^{\phantom{x,y}z} = \delta(xy,z) .
\end{align}
Here, the contraction of repeated indices is understood 
to represent the integration with the normalized Haar measure 
$\int d x \equiv \frac{1}{|G|} \sum_x$: 
\begin{align}
 y_{x,y}^{\phantom{x,y}z} e_{z}
 \equiv \int\! d z \,y_{x,y}^{\phantom{x,y}z} e_{z} 
 = \frac{1}{|G|} \sum_z y_{x,y}^{\phantom{x,y}z} e_{z} ,
\end{align}
and $\delta(x,y)$ is  the delta function with respect to this measure:
\begin{align}
 \delta(x,y) \equiv |G| \,\delta_{x,y}  , 
 \qquad \int\! dx\, f(x) \,\delta(x,y) = f(y).
\end{align}
From the definition we obtain
\begin{align}
 y_{x_1, x_2,\ldots, x_k} &= \delta (x_1 x_2 \cdots x_k,1), \\
g^{x,y} &= \delta(xy,1) ,
\end{align}
where $1$ is the identity of $G$.
Therefore, the action \eqref{hm2_action} can be written 
with the symmetric dynamical variables 
$A_{x,y} = A_{y,x}$ and $B^{x,y} = B^{y,x} $ 
as
\begin{align}
 S[A,B] &= \frac{1}{2} A_{x,y}\, B^{x,y}
 - \frac{\lambda}{6} A_{x^{-1},y}\, A_{y^{-1}, z}\, A_{z^{-1}, x} 
\nn\\
 &~~~- \sum_{k \geq 2} \frac{\mu_k}{2k}\, B^{x_1 y_1} \cdots
 B^{x_k y_k}\, \delta(x_1 \cdots x_k, 1)\, \delta(y_k \cdots y_1, 1).
\label{hm2gr_action1}
\end{align}

\subsection{The Feynman rules and the free energy for a group ring} 
\label{hm2gr_Feynman_rule}

The action \eqref{hm2gr_action1} can be rewritten 
to a form similar to that of matrix ring, 
by expressing everything in terms of the irreducible representations of $G$.   
To show this, we first write the delta function as
\begin{align}
\delta(x_1 \cdots x_k,1) = \sum_R d_R\, \mathrm{tr}(D_R(x_1) \cdots D_R(x_k)) ,
\end{align}
where the sum is taken over all the irreducible representations $R$ of $G$ 
with the representation matrix $D_R(x)=(D^R_{ab}(x))$ $(x\in G)$, 
and $d_R$ is  the dimension of representation $R$, 
$d_R=\mathrm{tr}\, D_R(1)$.
Then, the action \eqref{hm2gr_action1} can be rewritten to the form
\begin{align}
 S &= \frac{1}{2} \sum_{R,S} d_{R}\,d_{S} A^{RS}_{abcd}B^{RS}_{abcd} 
 - \frac{\lambda}{6}\sum_{R_1,R_2,R_3}
 d_{R_1} d_{R_2} d_{R_3}  A^{R_1R_2}_{a_1b_1b_2a_2}
 A^{R_2R_3}_{a_2b_2b_3a_3} A^{R_3R_1}_{a_3b_3b_1a_1} 
\nonumber 
\\
& \quad- \sum_{k \geq 2} \frac{\mu_k}{2k}
 \sum_{R,S} d_R \, d_S \, B^{RS}_{a_1a_2b_2b_1}
 \cdots B^{RS}_{a_ka_1b_1b_k} ,
\label{hm2_gr_action2} 
\end{align}
where
\begin{align}
 A^{RS}_{abcd} &\equiv \int \! dxdy \,A_{x,y}D^{R}_{ab}(x) D^{S}_{cd}(y)
 =  A^{SR}_{cdab} ,
\\
 B^{RS}_{abcd} &\equiv \int \! dxdy \,B^{x,y}D^{R}_{ba}(x^{-1})D^{S}_{dc}(y^{-1})
 = B^{SR}_{cdab}.  
\end{align}
This action gives the following Feynman rules:
\begin{align}
{\rm propagator}&:\ 
 \begin{array}{l} \includegraphics[height = 1.5cm]{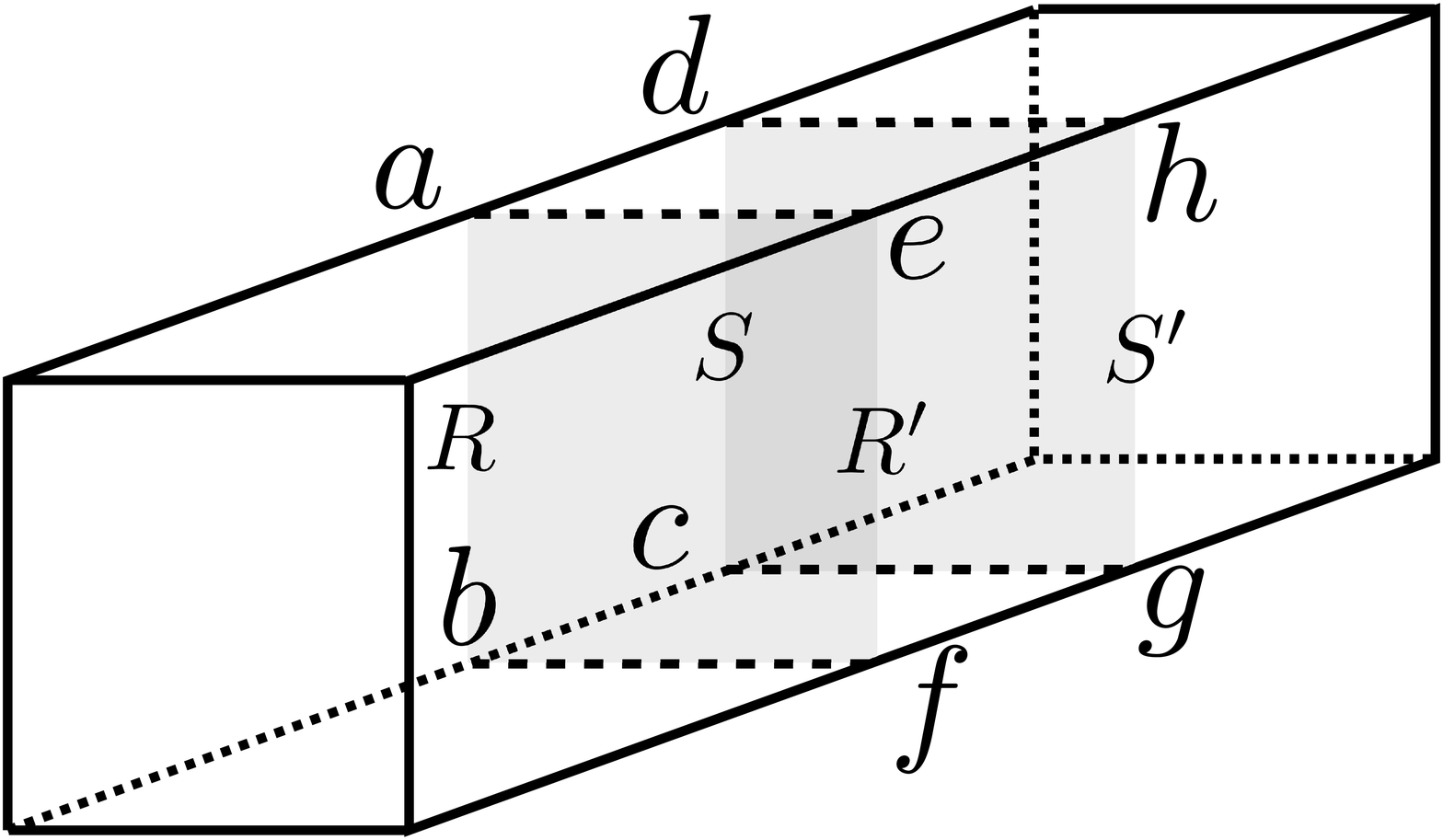}
 \end{array}
 + \begin{array}{l} \includegraphics[height = 1.5cm]{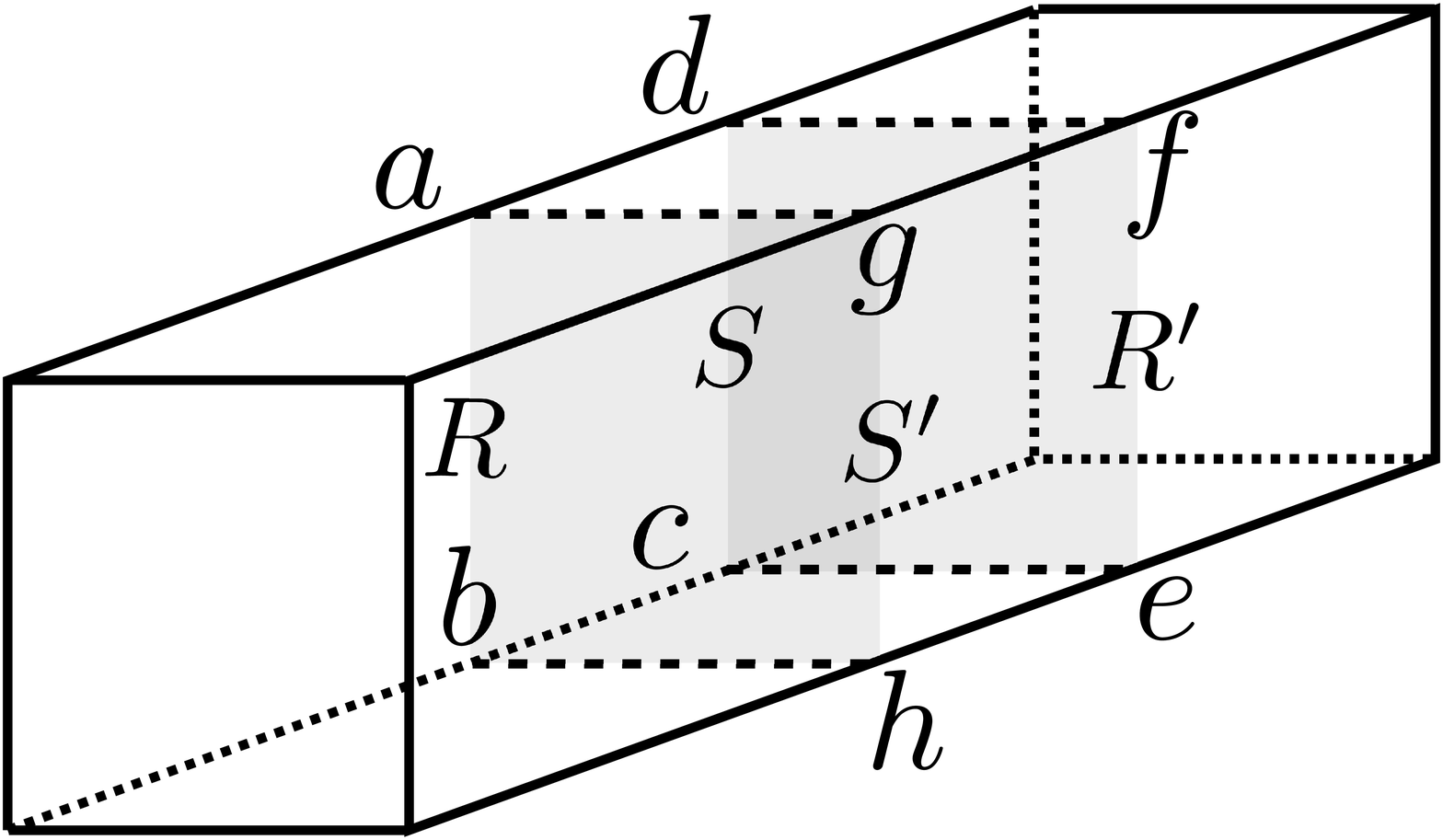}
 \end{array} 
\nonumber 
\\
 & \quad \sim \langle A^{RS}_{abcd}B^{R^\prime S^\prime}_{efgh} \rangle 
 = \frac{1}{d_R d_S}(\delta_{ae} \delta_{bf} \delta_{cg}
 \delta_{dh} \delta^{RR^\prime} \delta^{SS^\prime} 
 +  \delta_{ag} \delta_{bh} \delta_{ce} \delta_{df} \delta^{RS^\prime}
 \delta^{SR^\prime}) ,
\label{hm2gr_propagator} 
\\
 {\rm triangle}&:\ \begin{array}{l} \includegraphics[width = 4cm]{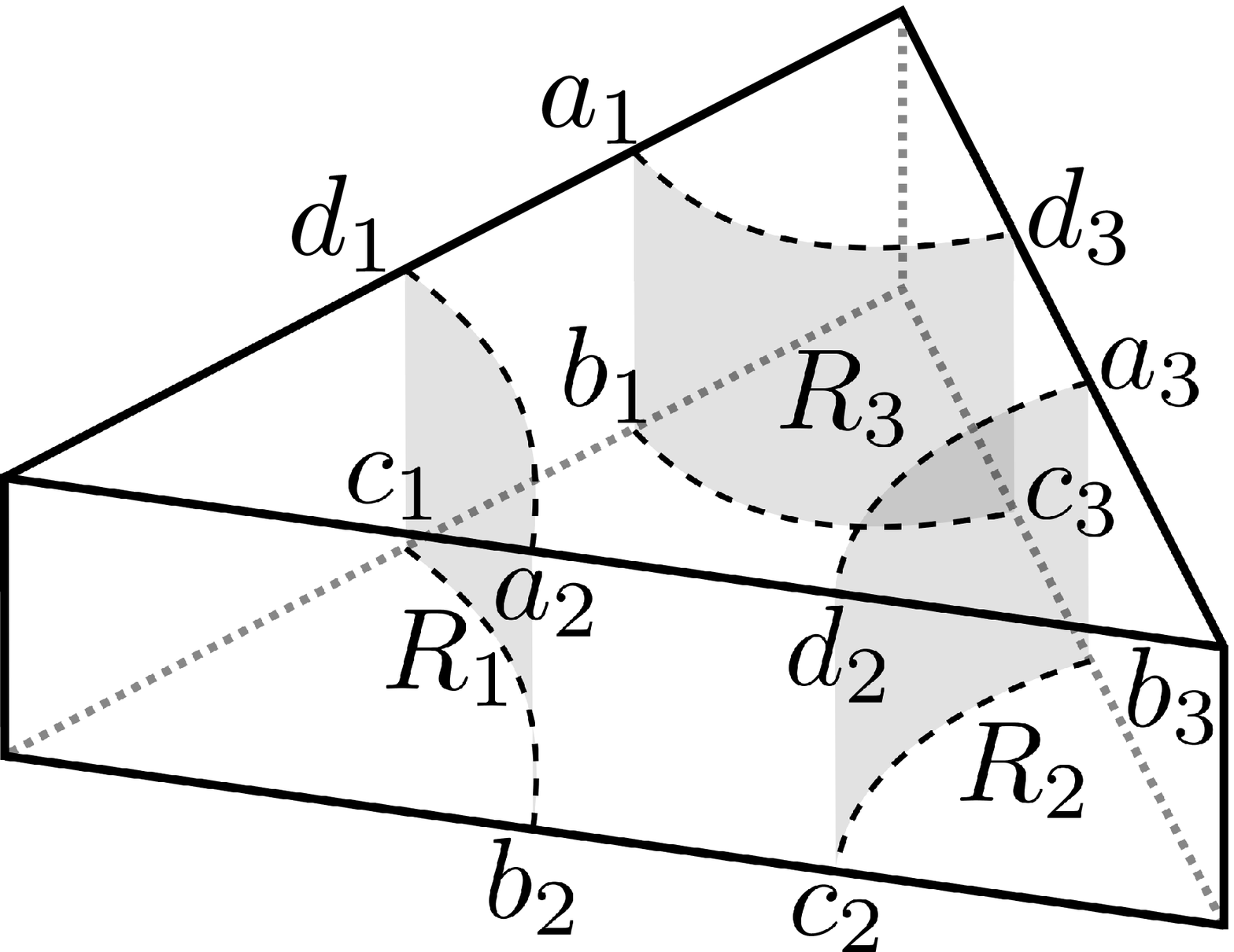}
 \end{array} 
 \sim \lambda\, d_{R_1} d_{R_2} d_{R_3} \delta_{a_1d_2} \delta_{c_1b_2}
 \ldots \delta_{a_kd_1} \delta_{c_kd_1} ,
\label{hm2gr_triangle} 
\\
 \mbox{$k$-hinge}&:\ \begin{array}{l} \includegraphics[width = 3cm]{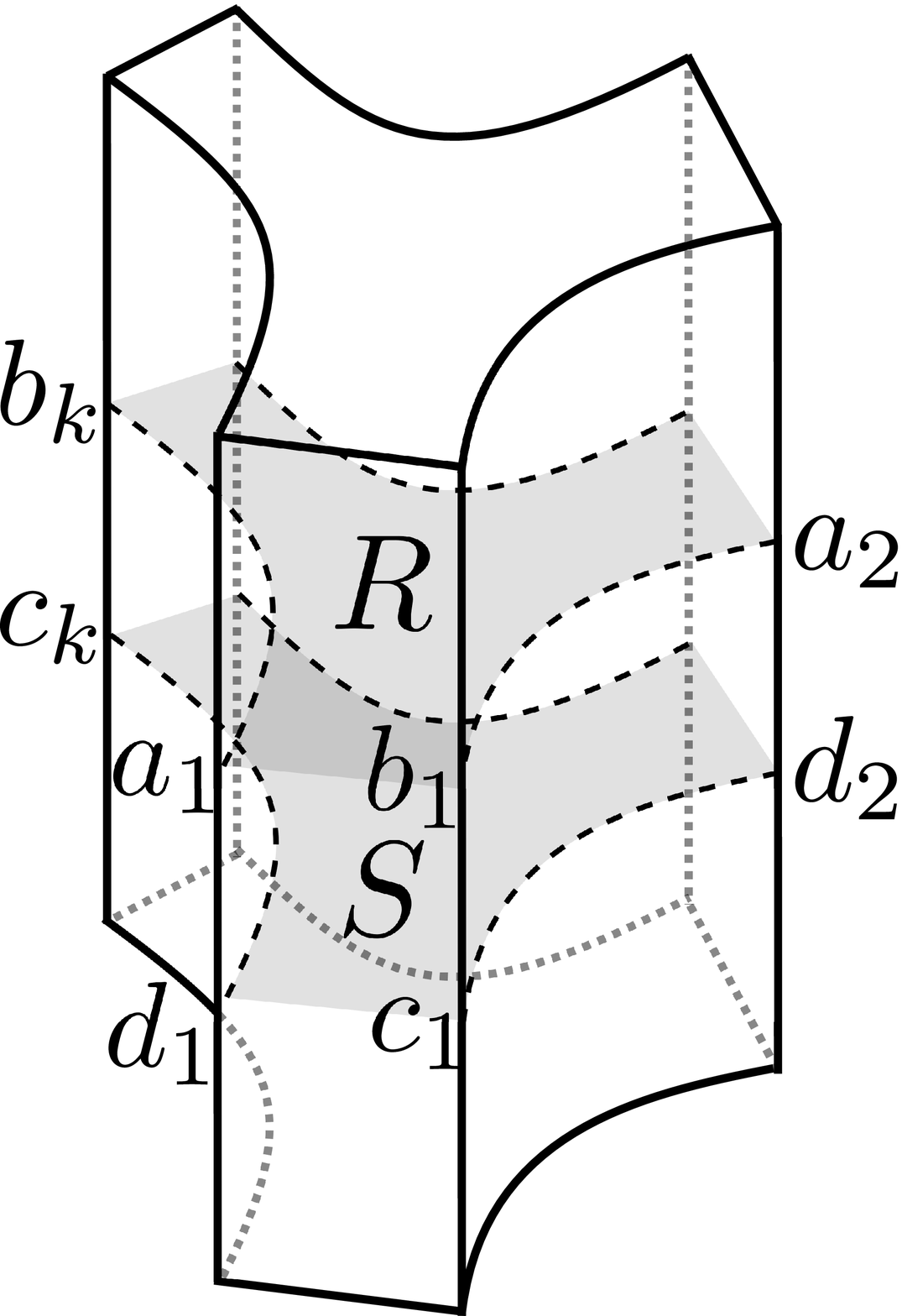}
 \end{array} 
 \sim \mu_k\, d_R\, d_S\, \delta_{a_1 b_2} \delta_{d_1 c_2} \ldots
 \delta_{a_k b_1} \delta_{d_k c_1} .
\label{hm2gr_hinge}
\end{align}
We thus see that the index network around every vertex 
is again expressed as a closed surface 
with double index lines, 
and its index function is determined 
only by the Euler characteristics of the  polygonal decomposition:%
\footnote{
This expression can be naturally understood 
if $\mathcal{F}(\gamma)$ is regarded as the real sector 
of the index function for the complexified algebra 
$\mathcal{A}^{\mathbb{C}}=\mathbb{C}[G]$, 
because the group ring $\mathbb{C}[G]$ 
can be expressed as the direct sum of $M_{d_R}(\mathbb{C})$ over $R$, 
$\mathbb{C}[G] = \bigoplus_R M_{d_R}(\mathbb{C})$. 
} 
\begin{align}
 \mathcal{F}(\gamma)
 = \prod_{v:\,\text{vertex}} \mathcal{I}_{g(v)}
 = \prod_{v:\,\text{vertex}} \Bigl[\sum_R (d_R)^{2-2g(v)}\Bigr] .
\end{align}
Here, elementary group theory  
shows that $\sum_R d_R^2=|G|$, 
and $\sum_R d_R^0$ gives the number of irreducible representations 
which equals that of conjugate classes. 
For example, when $G$ is the  symmetric group $S_n$, 
we have
\begin{align}
 \sum_R d_R^2 = |G| = n!\,,  \qquad \sum_R d_R^0 =p_n,
\label{zeta_group}
\end{align}
where $p_n$ denotes the number of partitions of $n$. 
Therefore, 
if $G$ admits the relations $\sum_R d_R^2 \gg \sum_R d_R^{2-2g}$ $(g\geq 1)$, 
the index networks of spherical topology 
have a large value of index function 
compared to those of higher genera.

We also can introduce a color structure as in subsection \ref{intro_color} 
and can control the number of vertices 
by considering the direct sum of $K$ copies of group ring 
as in subsection \ref{center}.
Therefore, we can again single out the diagrams 
which give tetrahedral decompositions of three-dimensional manifolds.

\section{Summary and discussion}
\label{conclusion}

In this paper we construct a class of models 
that generate random diagrams consisting of triangles and multiple hinges.  
The models are completely characterized 
by semisimple associative algebras $\mathcal{A}$ and tensors $C^{ijklmn}$. 
When $C^{ijklmn}$ are chosen as in \eqref{C_def} or \eqref{C_color}, 
each Feynman diagram can be expressed 
as a collection of index networks around vertices. 
The contribution $\mathcal{F}(\gamma)$ from each diagram $\gamma$ 
to the free energy 
is expressed as the product of 
the index functions $\zeta(v)$ of vertices $v$ of $\gamma$, 
and $\zeta(v)$ depends only on the topology of the index network around $v$ 
besides the structure of the defining associative algebra.

Although most of the Feynman diagrams do not represent 
three-dimensional manifolds, 
we give a general prescription to automatically reduce the set of possible diagrams 
such that only (and all of the) manifolds are generated.  
We implement the strategy for the models with $\mathcal{A}$ set to matrix rings, 
by introducing a color structure and taking the direct sum of $K$ copies of matrix ring. 
We show that every diagram actually gives a tetrahedral decomposition 
where each vertex has a neighborhood homeomorphic to $B^3$ 
(ensured by the statement that the index network around each vertex 
has the topology of $S^2$). 

We further demonstrate that there is a novel strong-weak duality in the models 
which interchanges the roles of triangles and hinges. 
We also investigate the models 
where the defining associative algebras are group rings, 
and show that most of  their analytic properties can be understood 
as a straightforward generalization of those for matrix rings. 

We now list some of the future directions for further study of the models. 
The first is about the topology summation. 
Our models actually give a summation over all topologies of three-dimensional manifolds. 
It seems that we cannot distinguish the topology of the Feynman diagrams 
if the tensor $C^{ijklmn}$ has the form 
\eqref{C_def} or \eqref{C_color} as we took in this paper, 
because topologically different diagrams can give contributions 
of the same form to the free energy. 
One can optimistically think that this represents membrane instability 
(see, e.g., \cite{Taylor:2001vb}).   
However, it may also happen that configurations of some specific topology 
entropically dominate in a critical region,   
although we have not fully evaluated the numerical coefficients in the free energy 
and their dependence on topology.  

Another way to investigate the topologies of diagrams 
is to change the tensor $C^{ijklmn}$ to other forms. 
In fact, this change significantly modifies 
the dependence of the index function on the associative algebra $\mathcal{A}$.  
The change of $C^{ijklmn}$ and its effect on topological invariants 
will be studied in our future paper \cite{fsu2}.

The second direction for further study is about the continuum limit. 
Although we have not fully studied the continuum limit yet,%
\footnote{
Note that our models become topological 
when we set $\mu_k N=1$ and $\lambda^2 \mu_k=1$ $(k\geq 2)$,  
since the dependence of $s_0$ and $s_3$ disappear from \eqref{wt_general} 
[or \eqref{free_energy_FSU}]. 
These values of coupling constants may correspond 
to a certain (possibly uninteresting) critical point 
because the models are not only diffeomorphism invariant 
but also Weyl invariant. 
} 
there may be a chance to analytically solve the models in the large $N$ limit 
and to determine the critical behaviors, 
because the dynamical variables of our models 
are given by symmetric matrices $A_{ij}$ and $B^{ij}$.

The third direction is about the introduction of matters to our models. 
We expect that extra degrees of freedom representing matter fields 
can be introduced 
by making copies of the variables $A_{ij}$ and $B^{ij}$ 
as in matrix models \cite{Kazakov:1988ch}. 
It should be particularly interesting to introduce matter fields 
corresponding to the target space coordinates 
of embedded membranes and to study the critical behaviors.
It would then be important to investigate 
if there is an analogue of the so-called ``$c=1$ barrier'' in the models 
and how the situation is modified when supersymmetry is introduced.

We close this section with a brief comment on the relationship 
of our models with the colored tensor models. 
It is worth noting that our models (with a color structure 
and an appropriate limit of parameters as in subsection \ref{tetrahedron}) 
generate {\em all}\, of the possible tetrahedral decompositions of three-dimensional 
manifolds, 
and thus should have more configurations than those of the colored tensor models.  
For example, the colored tensor models 
do not generate such tetrahedral decompositions 
where odd number of triangles are glued together along a hinge. 
Since the colored tensor models introduce a pair of tensors as in two-matrix models,  
they may correspond to three-dimensional gravity 
with specific matters. 
In this sense, our models with minimum fine tunings 
may give a continuum theory (if exists) closer to pure gravity.%
\footnote{
Of course, it is highly possible that the two models 
are in the same universality class defining pure gravity.
} 
It should be interesting to investigate 
whether colored tensor models can be obtained 
by adding some (not necessarily unitary) matters to our models.

In a remarkable paper \cite{Bonzom:2012hw}, 
it is shown that the free energy of three-dimensional colored tensor models 
depend on the size of tensor,  $\mathcal{N}$, as
\begin{align}
\mathcal{N}^3\sum_{\mathcal{G}}\mathcal{N}^{-\omega(\mathcal{G})}, 
\end{align}
where we have suppressed other coupling constants. 
$\mathcal{G}$ denotes a colored graph 
which is dual to a tetrahedral decomposition of three-dimensional pseudomanifold, 
and $\omega(\mathcal{G})$ is the degree of $\mathcal{G}$ 
(see \cite{Bonzom:2012hw} for details).  
Thus, in the large $\mathcal{N}$ limit, the leading contribution comes 
from colored graphs with $\omega(\mathcal{G})=0$.   
It is also shown that if $\omega(\mathcal{G})=0$ then $\mathcal{G}$ 
is dual to a three-sphere \cite{Gurau:2011xq}. 
Therefore, the leading order graphs are homeomorphic to $S^3$.  
We can say the same thing for our models
if we confine our attention to 
only the diagrams that have tetrahedral decompositions dual to colored graphs.  
The degree $\omega$ of such a diagram can be evaluated as in \cite{Bonzom:2012hw} 
and becomes 
\begin{align}
 \omega =\frac{3}{2}s_3 +3-s_1,  
\end{align}
with which the contribution \eqref{free_energy_FSU} to the free energy 
can be rewritten to the form 
\begin{align}
 \frac{1}{S} \Bigl(\frac{N}{\lambda^2}\Bigr)^{-\frac{2}{3}\omega +2} 
 \bigl(\lambda^4 \mu^3 N \bigr)^{\frac{1}{3} s_1} .
\end{align}
Thus, if we take a limit $N/\lambda^2 \to \infty$  
with $\lambda^4 \mu^3 N$ kept finite, 
the leading contribution comes from configurations with $\omega=0$, 
that is, tetrahedral decompositions of $S^3$.   
It is interesting to study the meaning of the degree 
for general tetrahedral decompositions which are not dual to colored graphs.

\section*{Acknowledgments}
The authors thank Hikaru Kawai for stimulating discussions. 
MF is supported by MEXT (Grant No.\,23540304).
SS is supported by the JSPS fellowship.

\baselineskip=0.83\normalbaselineskip


\end{document}